\PassOptionsToPackage{dvipsnames}{xcolor}
\documentclass[times]{aastex631}

\usepackage{amsmath}
\usepackage{CJK}
\usepackage{bm}
\usepackage[normalem]{ulem}

\makeatletter
\newcommand*{\rom}[1]{\expandafter\@slowromancap\romannumeral #1@}
\makeatother

\usepackage{dcolumn}
\newcolumntype{q}[1]{D{.}{.}{#1}}

\def\dfvmwdv{DAVE4VM\-wDV}

\usepackage{newunicodechar,graphicx}
\DeclareRobustCommand{\okina}{%
  \raisebox{\dimexpr\fontcharht\font`A-\height}{%
    \scalebox{0.8}{`}%
  }%
}

\newcommand{\chg}[1]{#1}

\begin{document}

\begin{CJK*}{UTF8}{gbsn}
\title{What Can \textit{DKIST}/DL-NIRSP Tell Us About Quiet-Sun Magnetism?}

\author[0000-0002-7290-0863]{Jiayi Liu (刘嘉奕)}
\altaffiliation{\textit{DKIST} Ambassador}
\affiliation{Institute for Astronomy, University of Hawai$\okina$i at M\={a}noa, 2680 Woodlawn Dr., Honolulu, HI 96822, USA}

\author[0000-0003-4043-616X]{Xudong Sun (孙旭东)}
\affiliation{Institute for Astronomy, University of Hawai$\okina$i at M\={a}noa, 34 Ohia Ku Street, Pukalani, HI 96768, USA}

\author[0000-0003-1522-4632]{Peter W. Schuck}
\affiliation{Heliophysics Science Division, NASA Goddard Space Flight Center, 8800 Greenbelt Rd., Greenbelt, MD 20771, USA}

\author[0000-0001-5459-2628]{Sarah A. Jaeggli}
\affiliation{National Solar Observatory, 22 Ohia Ku Street, Pukalani, HI 96768, USA}

\correspondingauthor{Jiayi Liu (刘嘉奕)}
\email{jiayiliu@hawaii.edu}

\begin{abstract}

Quiet-Sun regions cover most of the Sun's surface; its magnetic fields contribute significantly to the solar chromospheric and coronal heating. However, characterizing the magnetic fields of the quiet Sun is challenging due to their weak polarization signal. The 4-m \textit{Daniel K. Inouye Solar Telescope} (\textit{DKIST}) is expected to improve our understanding of the quiet-Sun magnetism. In this paper, we assess the diagnostic capability of the Diffraction-Limited Near Infrared Spectropolarimeter (DL-NIRSP) instrument on \textit{DKIST} on the energy transport processes in the quiet-Sun photosphere. To this end, we synthesize high-resolution, high-cadence Stokes profiles of the \ion{Fe}{1} 630~nm lines using a realistic magnetohydrodynamic simulation, degrade them to emulate the \textit{DKIST}/DL-NIRSP observations, and subsequently infer the vector magnetic and velocity fields. For the assessment, we first verify that a widely used flow-tracking algorithm, Differential Affine Velocity Estimator for Vector Magnetograms, works well for estimating the large-scale ($> 200$ km) photospheric velocity fields with these high-resolution data. We then examine how the accuracy of inferred velocity depends on the temporal resolution. Finally, we investigate the reliability of the Poynting flux estimate and its dependence on the model assumptions. The results suggest that the unsigned Poynting flux, estimated with existing schemes, can account for about $71.4\%$ and $52.6\%$ of the reference ground truth at $\log \tau =0.0$  and $\log \tau = -1$. However, the net Poynting flux tends to be significantly underestimated. The error mainly arises from the underestimated contribution of the horizontal motion. We discuss the implications on \textit{DKIST} observations.

\end{abstract}



\section{Introduction} \label{sec:intro}
The quiet Sun refers to the region outside the sunspots and active regions, which occupies most of the solar surface at all times. The magnetic fields are organized as a complex network entrained between the convective cells. While the mean magnetic flux density is low, the total unsigned magnetic flux is comparable to that of the active regions, and the flux emergence rate is larger than that of the active regions \citep{QSflux,QSflux_rate}. 

The quiet Sun magnetic fields are important in contributing to the energy budget in the solar atmosphere. New and existing magnetic fields frequently interact with each other and with convective flows, which can lead to ubiquitous nanoflares and magnetohydrodynamic (MHD) waves \citep{Jess2023}. MHD simulations of the quiet Sun suggest that these interactions can provide enough energy to heat the chromosphere and corona \citep{Rempel2014,QSenergy}. Recent observations from \textit{Solar Orbiter} \citep{campfire} indicate that solar atmosphere above the quiet Sun is in fact very dynamic at small scales. 

\chg{Since the discovery of sunspot magnetic fields \citep{Hale}, a lot of effort has been made to extract information from the solar spectra, for example, by inverting the radiative transfer equation \citep{inversion} or by using radio methods \citep{radio}. Spectropolarimetry, which measures wavelength-dependent Stokes parameters $(I,Q,U,V)$ in magnetically sensitive spectral lines, is one of the powerful methods of diagnosing the solar magnetic field.} The shape of Stokes profiles contains information about temperature, magnetic field, and other MHD state variables of the solar atmosphere. These parameters can be inferred by solving the polarized radiative transfer equations, a process known as ``inversion."

The quiet-Sun magnetic fields are challenging to infer because of their relatively low mean flux density and thus weak polarization signal \citep{QSreview}. The study of quiet-Sun energy transport is further limited by the spatial and temporal resolution of observations. According to MHD simulations of the quiet Sun, 50\% of the energy resides on a scale smaller than 100 km. A sampling rate of $8$~km or smaller is needed to properly recover the spectral energy distribution \citep{Rempel2014}, which requires 10-m-class telescope aperture at the optical wavelengths. For the typical quiet Sun, slit-based spectrographs often require long integration time and thus tens of minutes to assemble a raster. The average lifetime of internetwork magnetic elements, however, is only about ten minutes \citep{deWijn2008,Zhou2010}. The quiet-Sun magnetic fields will have significantly evolved during the scan.  

Despite these difficulties, efforts have been made in estimating the energy flux through quiet-Sun photosphere. Use observations from the \textit{Hinode} satellite, \cite{Giannattasio} studied the supergranular spatial and temporal scales and found that the energy flux is sufficient in sustaining the magnetic fields in the network. In contrast, \cite{Tilipman} found the energy flux inferred from the SUNRISE mission data difficult to match the radiative loss of solar corona at the granular scales. 

The 4-m \textit{Daniel K. Inouye Solar Telescope} \citep[\textit{DKIST};][]{DKIST} is expected to improve our understanding of small-scale magnetism and energy transport via its high-resolution spectropolarimetric observations \citep{DKSIT_PLAN}. The Diffraction-Limited Near-Infrared Spectropolarimeter \citep[DL-NIRSP;][]{DLNIRSP}, one of the first-light instruments of DKIST, can measure Stokes parameters at an effective resolution of 0.03$''$ ($\sim$22~km) with high spectral resolution and polarimetric accuracy. Using an integral field unit, it can obtain spectropolarimetric information of a two-dimensional (2D) field of view (FOV) at the same time, at a fast cadence of several seconds.

In this paper, we will investigate how well DL-NIRSP observations can characterize the energy transport in the quiet Sun. We will use a realistic MHD simulation to create synthetic DL-NIRSP observations for the solar photosphere. Subsequently, we will use an inversion algorithm to infer the depth-dependent vector magnetic fields, and finally use these magnetic maps to estimate the velocity fields and the Poynting flux. The derived quantities will be compared with the MHD ground truth. 

The rest of the paper is organized as follows. In section \ref{sec:Data}, we describe the MHD model, the Stokes synthesis and inversion algorithm, and the flow tracking algorithm for velocity estimation. In section \ref{sec:DAVE4VMwDV}, we focus on validating the flow tracking method on high resolution data. In section \ref{sec:synthesis}, we present the results of estimating energy transport from the emulated observation data. In section \ref{sec:discussion}, we discuss our results and analyze the possible causes for discrepancies. In section \ref{sec:conclusion}, we draw our conclusions. 


\section{Data and Method} \label{sec:Data}
\subsection{MHD Simulation}

To simulate DL-NIRSP's high resolution observations, we use the MPS/University of Chicago Radiative MHD \citep[MURaM;][]{MURaM} simulation, which is widely used in solar dynamo studies. The MURaM code solves realistic equations of state with partial ionization and radiative transfer in three dimensions (3D) to simulate the relevant atmospheric parameters. The simulations can produce small-scale magnetic features that well resemble the high resolution observations of solar photosphere \citep{obs}. The model atmospheres are calculated in a Cartesian coordinate frame.

The model used here is from the quiet-Sun magnetoconvection run \texttt{O16bM} from \cite{Rempel2014}. It is initialized with a distribution of mixed polarity fields with an average magnetic flux density of $\langle B \rangle \sim 120$~G at $\log \tau_{500} = 0$, where $\tau_{500}$ represents the optical depth at continuum wavelength $500$~{nm}. Hereafter we use $\tau$ to represent $\tau_{500}$. The computation box contains $1536\times 1536\times 128$ pixels with a \chg{pixel size} of 16 km in the horizontal direction and 12 km in the vertical direction. The temporal step is $2$~s. These high-resolution simulations were also used to understand the instrument influences for \textit{Hinode}/SP \citep{hinode} and \textit{SUNRISE}/IMaX \citep{imax}.

\begin{figure*}[t!]
  \centering
  \includegraphics[width=0.98\textwidth]{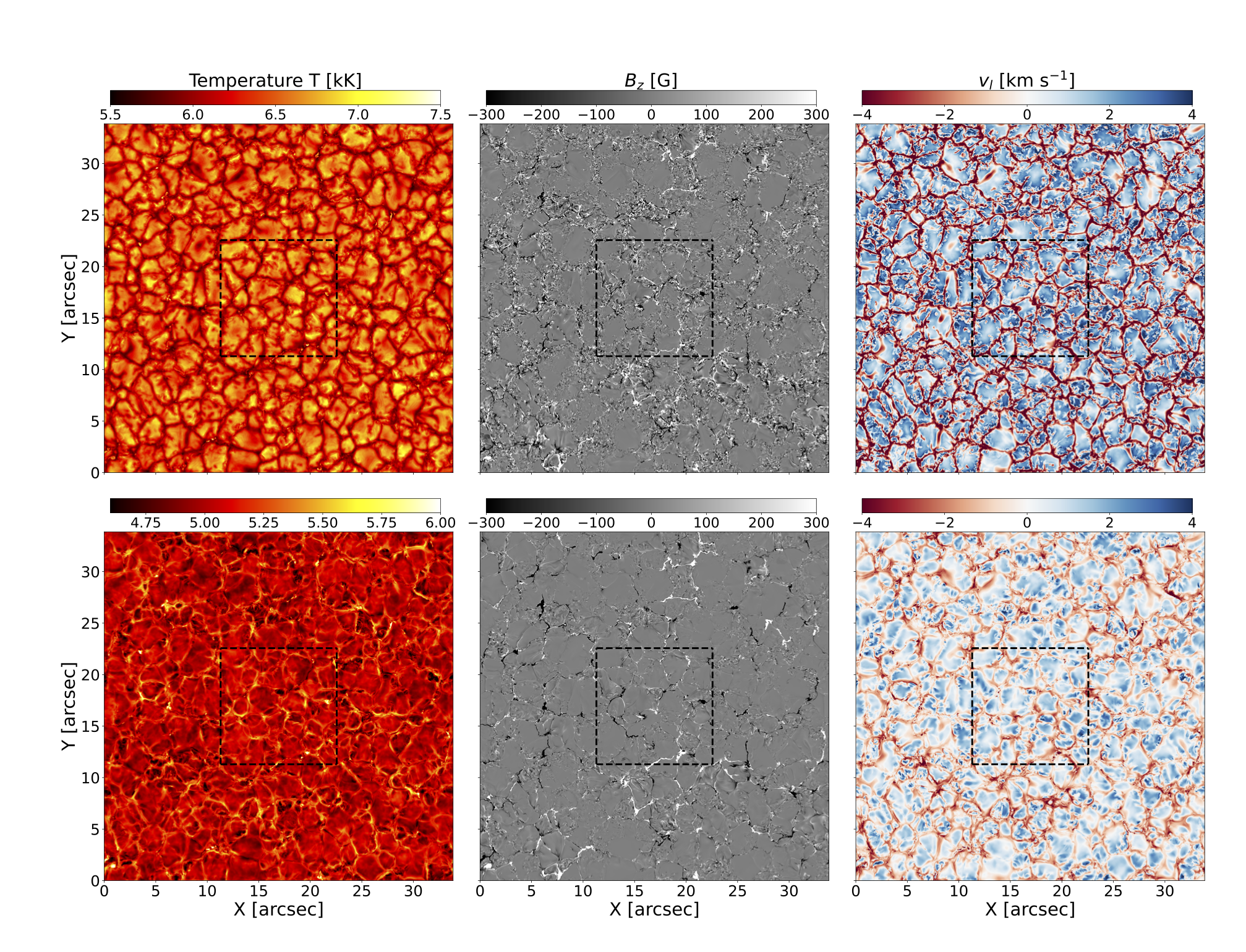}
  \caption{Overview of MURaM simulation at $\log \tau = 0$ (top) and $\log \tau = -1$ (bottom) at $t=2$~s. From left to right, the maps are for temperature $T$, vertical magnetic field $B_z$, and vertical velocity $v_l$. The black dashed box marks the region of interest in this work.}
  \label{fig:MURaM}
\end{figure*}

In this paper, we analyze MURaM simulation with a cadence of $\Delta t = 2$~s, which is the highest cadence made available to us. We choose a region of interest of $512 \times 512 \times 128$ pixels from the center of the simulation box. For each frame, we make use of the following MHD variables: temperature $T$, gas density $\rho$, gas pressure $P_g$, electron pressure $P_e$, vector velocity field $\bm{v}$, and vector magnetic field $\bm{B}$. An example of the temperature, vertical field $B_z$, and vertical velocity $v_z$ map at $t = 2$~s at $\log \tau = 0.0$ and $-1.0$ are shown in Figure \ref{fig:MURaM}. Vigorous convective patterns are more pronounced at $\log \tau = 0$ compared to $\log \tau = -1$. The root mean square magnetic field ($B_\mathrm{rms}$) is $198.4$~G at $\log \tau = 0$ compared to $B_\mathrm{rms}$ equal to $129.3$~G at $\log \tau = -1$.

\subsection{Stokes Synthesis and Inversion}

Many inversion methods have been developed to extract information from the Stokes observations \citep{inversion}. One widely used algorithm is the Stokes Inversion based on Response functions \citep[SIR;][]{SIR}, which is capable of inferring depth-dependent physical properties from multiple spectral lines. Given a set of MHD state variables along the line of sight (LOS), SIR can also forward model the emergent Stokes profiles. SIR operates under the assumption of hydrostatic equilibrium and local thermodynamic equilibrium (LTE). 

In the synthesis mode, SIR numerically solves the polarized radiative transfer equation in magnetically sensitive lines
\begin{equation}
\frac{d \bm{I}(\tau)}{d \tau} = K(\tau)\left[\bm{I}(\tau) - \bm{S}(\tau)\right],
\label{equ:di5dt}
\end{equation}
where $\bm{I}=(I,Q,U,V)$ is the Stokes vector with polarization signal induced by the Zeeman effect, $K$ is the total absorption matrix, $\bm{S} = (S_I, S_Q, S_U, S_V) = (B_\nu (T),0,0,0)$ is the source function vector, and $B_\nu(T)$ is the Planck function. Here $K$ and $\bm{S}$ are evaluated under the assumption of local thermal equilibrium (LTE). To simplify the problem, we synthesize four Stokes profiles only at disk center with a spectral sampling rate of $8.95$~m{\AA} over a spectral window of $2.459$~{\AA}, which covers from $-655.9$ to $1803.9$~m{\AA} from the rest wavelength of the line core of \ion{Fe}{1} at 630.15~nm. These parameters are typical for \textit{DKIST}/DL-NIRSP observations. Furthermore, we convolve the synthesized spectra with the theoretical point spread function of \textit{DKIST}/DL-NIRSP and resample the pixel size to $0.03\arcsec$ (about $22$~km) to emulate a DL-NIRSP observation using High-Res mode ($f/62$) which has a resolution at diffraction limit \citep[$0.03\arcsec$;][]{DLNIRSP}. We did not add noise to the synthetic spectra for this study. The effect of noise can be important \citep[e.g.,][]{quiteronoda2023}, however we will defer its investigation to future work.

In the inversion mode, SIR modifies the initial model atmosphere iteratively until the synthetic Stokes profiles match the observed ones. It returns the $\tau$-dependent temperature, LOS velocity, magnetic field strength, inclination, and azimuth angles along the LOS. The fitting is performed by introducing perturbations to the initial guess of the atmosphere, and minimizing the $\chi^2$ between the observed and modeled Stokes profiles
\begin{equation}
  \chi^2 = \frac{1}{\nu}\sum_{k=1}^{4}\sum_{i=1}^{M} \left[I_k^\text{{obs}}(\lambda_i)-I_k^{\text{syn}}(\lambda_i)\right]^2\frac{\omega_{k}^2}{\sigma_{k}^2},
  \label{equ:SIR_chi2}
\end{equation}
where $\nu$ is the number of free parameters, $k$ is the index of the four Stokes parameters, $i$ is the index of the wavelength points, and $\omega_{k}$ and $\sigma_{k}$ are the weights and uncertainties for each Stokes parameter.
In order to reduce the number of free parameters, the parameters are evaluated only at several ``nodes", i.e., grid points with fixed $\tau$, and the values at the remaining $\tau$ grid points are approximated by interpolation. To extract the information from degraded Stokes profiles, we invert the spectra with the node configuration listed in Table \ref{tab:configuration} \citep{quiteronoda2023}. The weights for Stokes $(I, Q, U, V)$ in chi-square evaluation are $(2, 20, 20, 5)$, respectively. 

We note that the inferred variables are typically a function of $\tau$. A 2D map generally represents a slice with constant $\tau$ with varying geometric height $z$. Such is the limitation of many inversion algorithms (including SIR), which assumes hydrostatic equilibrium and lacks an absolute height scale along the LOS \citep{FIRTEZ-dz}. 

\begin{deluxetable}{ccccc}[t!]
  \tablecaption{Summary of SIR algorithm configuration}\label{tab:configuration}
  \tablewidth{0pt}
  \tablehead{
  \colhead{}&
    \multicolumn{4}{c}{{Nodes}} \\
  \cline{2-5}
  \colhead{\textbf{Parameters}} & \colhead{\textbf{Cycle 1}} & \colhead{\textbf{Cycle 2}} & \colhead{\textbf{Cycle 3}} & \colhead{\textbf{Cycle 4}} 
  }
  \startdata
    Temperature     & 2 & 3 & 5 & 5 \\ 
    Microturbulence & 1 & 1 & 1 & 1 \\ 
    LOS velocity    & 1 & 2 & 3 & 5\\ 
    Magnetic field strength  & 1 & 2 & 3 & 5 \\ 
    Inclination     & 1 & 2 & 3 & 5 \\ 
    Azimuth         & 1 & 2 & 2 & 2 \\ 
  \enddata
  \end{deluxetable}
\vspace{-5mm}


\subsection{Resolving 180-degree azimuthal ambiguity}

For the Zeeman effect, magnetic fields with azimuths differing by $180^\circ$ produce exactly the same linear polarization state. Additional assumptions are required to resolve this ambiguity \citep{metcalf, leka}. One widely used algorithm is the minimum energy method \citep{ME0}. It disambiguates the magnetic field by minimizing the summed absolute divergence of the magnetic field and the vertical electric current density. 
\begin{equation}
  E = \sum \left| \nabla \cdot \bm{B} \right| + \lambda \sum \left| J_z \right|
\end{equation}
where $\lambda$ is a weighting factor that controls the relative importance of the vertical electric current density. The vertical derivative in the divergence of magnetic field is obtained from a potential field calculated directly from the vertical magnetic field $B_z$.

In our study, we apply this algorithm to the magnetic fields inverted by SIR at $\log \tau = 0$ and $\log\tau = -1.0$, assuming the $\bm{B}$ vectors fall on a geometric height. As we shall see, the assumption is not strictly valid, but is a necessary simplifying step. We set $\lambda = 0.5$ as it gives a smooth solution. 


\subsection{Velocity Fields and Poynting Flux Estimate}

The temporal change of magnetic energy $E_m$ in a solar atmospheric volume $V$ with boundary $S$ can be calculated with the equation that involves the electric field $\bm{E}$, magnetic field $\bm{B}$, current density $\bm{J}$ and velocity field $\bm{v}$:

\begin{equation}
  \frac{dE_m}{dt} = \frac{1}{4\pi}\oint_S \bm{E} \times \bm{B} \cdot \bm{n} \, dS
    - \frac{1}{c}\int_V \bm{v} \cdot (\bm{J}\times \bm{B}) \, dV.
  \label{equ:dEdt}
\end{equation}
where $\bm{n}$ is the normal unit vector of the boundary. The first term on the right-hand side is the vertical Poynting flux ($S_z$), which measures the flow of electromagnetic energy through the boundary. In practice, we only consider the Poynting flux through the bottom boundary, which is usually the photosphere. The second term on the right-hand side is the power of the work done by the Lorentz force which converts kinetic energy to magnetic energy. The second term is usually ignored under the force-free condition. We will examine this approach in Section~\ref{subsec:etqs}.

The electric field $\bm{E}$ may be measured via the Stark effect or may be estimated by using the ideal Ohm's law ($\bm{E} = -\bm{v}\times \bm{B}$). The former method has been recognized to be critically affected by the low sensitivity of observations \citep{moran1991}. The latter method requires the knowledge of the full vector velocity field. \chg{While the LOS velocity $v_l$ can be directly estimated from the Doppler effect, the full vector velocity field $\bm{v}$ need to be estimated from flow-tracking methods \citep[e.g.,][]{welsch2007,schuck}.} For example, the velocity can be inferred by taking advantage of the physical relation between $\bm{v}$ and $\bm{B}$ via the induction equation \citep[e.g.,][]{Kusano2002,welsch2004, Longcope2004, schuck2006, schuck}, in particular the normal component of its ideal version:
\begin{equation}
    \frac{\partial{B_z}}{\partial t} = - \nabla_h \cdot (B_z\bm{v}_h-\bm{B}_hv_z),
  \label{equ:ideal}
\end{equation}
where the subscript $h$ denotes the horizontal component that is parallel to the photosphere. Given a time sequence of $\bm{B}$ maps, one may infer the velocity fields $\bm{v}$ using Equation~(\ref{equ:ideal}). The problem is generally not well posed and requires additional constraints. 

The Differential Affine Velocity Estimator for Vector Magnetograms \citep[DAVE4VM,][]{schuck} is a widely used, local flow tracking method. It estimates the plasma velocities by minimizing the L2 norm of the residual (difference of the left and right-hand side) of Equation~(\ref{equ:ideal}) within a windowed subregion. Recently, the algorithm has been modified to include the observed Doppler velocity $v_l$ as an additional constraint. Termed DAVE4VMwDV (with Doppler Velocity; P. W. Schuck 2024, in preparation), the global loss function $L$ now reads 
\begin{equation}
\begin{split}
L &= L_1 + \lambda L_2, \\ 
L_1 &= \sum_w \left\{ \frac{\partial B_z}{\partial t} + \bm\nabla _h \cdot (B_z \bm{v}_h-\bm{B}_hv_z)\right\}^2 \sigma^{-2}_{\partial_t B_z}, \\
L_2 &= \sum_w (\hat{\bm{\eta}}\cdot \bm{v}-v_l)^2 \sigma^{-2}_{v_l}.
\label{l2}
\end{split}
\end{equation}
where $\bm{B}$ and $v_l$ are the observational input, and $\bm{v}$ is the output. Here $\lambda$ is a scalar multiplier that controls the importance of the $L_2$ term, the subscript $w$ denotes the window, $\sigma_X$ refers to the uncertainty of variable $X$ ($\partial_t B_z$ or $v_l$), and the unit vector $\hat{\bm{\eta}}$ specifies the LOS direction. The operator $\bm{\nabla}_h$ acts on the horizontal components alone. In practice, we use a three-point stencil to calculate the time derivative, and a five-point stencil for the spatial derivatives, specifically,
\begin{equation}
  \begin{split}
    \left.\frac{\partial f}{\partial t}\right|_{t = t_0} &= \frac{f(t_0+\Delta t) - f(t_0 - \Delta t)}{2\Delta t}, \\  
  \label{diff}
  \end{split}
  \end{equation}
where we quote $\Delta t$ as the cadence of the output. We opt for this numerical scheme so that the inferred velocity can be co-temporal with the magnetogram: both are needed for the Poynting flux estimate. Appendix~\ref{app:successive} provides some discussion on this choice.

The algorithm requires several free parameters: the degree of Legendre expansion for horizontal velocity $d$ and vertical velocity $d_r$, the window size $w$, and relative weighting factor $\lambda$. These parameters can be empirically optimized to reduce $L$.


The inferred vector velocity field from DAVE4VMwDV then allows us to estimate the vertical Poynting flux $S_z$ passing through the photosphere via a surface integral
\begin{equation}\label{equ:Poynting}
\begin{split}
S_z & = S_z^\text{em} + S_z^\text{sh} \\
    & = \frac{1}{4\pi}\int_S B_h^2 v_z \, dS - \frac{1}{4\pi} \int_S (\bm{B}_h\cdot \bm{v}_h)B_z \, dS .
\end{split}
\end{equation}
Here the Poynting flux is typically divided into two terms. The emerging term $S_z^\text{em}$ corresponds to the first integral, which measures the energy transport due to the emergence (vertical transport) of magnetic flux tubes. The shearing term $S_z^\text{sh}$ corresponds to the second integral, which measures the energy transport resulting from the horizontal flow that shears the magnetic field.


\section{DAVE4VMwDV on Simulation Data} \label{sec:DAVE4VMwDV}

The DAVE4VM algorithm was originally tested on an MHD simulation with a $348$~km grid size \citep{schuck}. It has been applied widely on existing, low-resolution magnetograms with $\sim$ $1\arcsec$ resolution ($\sim720$~km), and mostly on active regions with stronger magnetic fields. Its performance in the high-resolution, quiet-Sun regions has not been demonstrated.


\begin{figure*}[t!]
  \centering
  \includegraphics[width=0.95\textwidth]{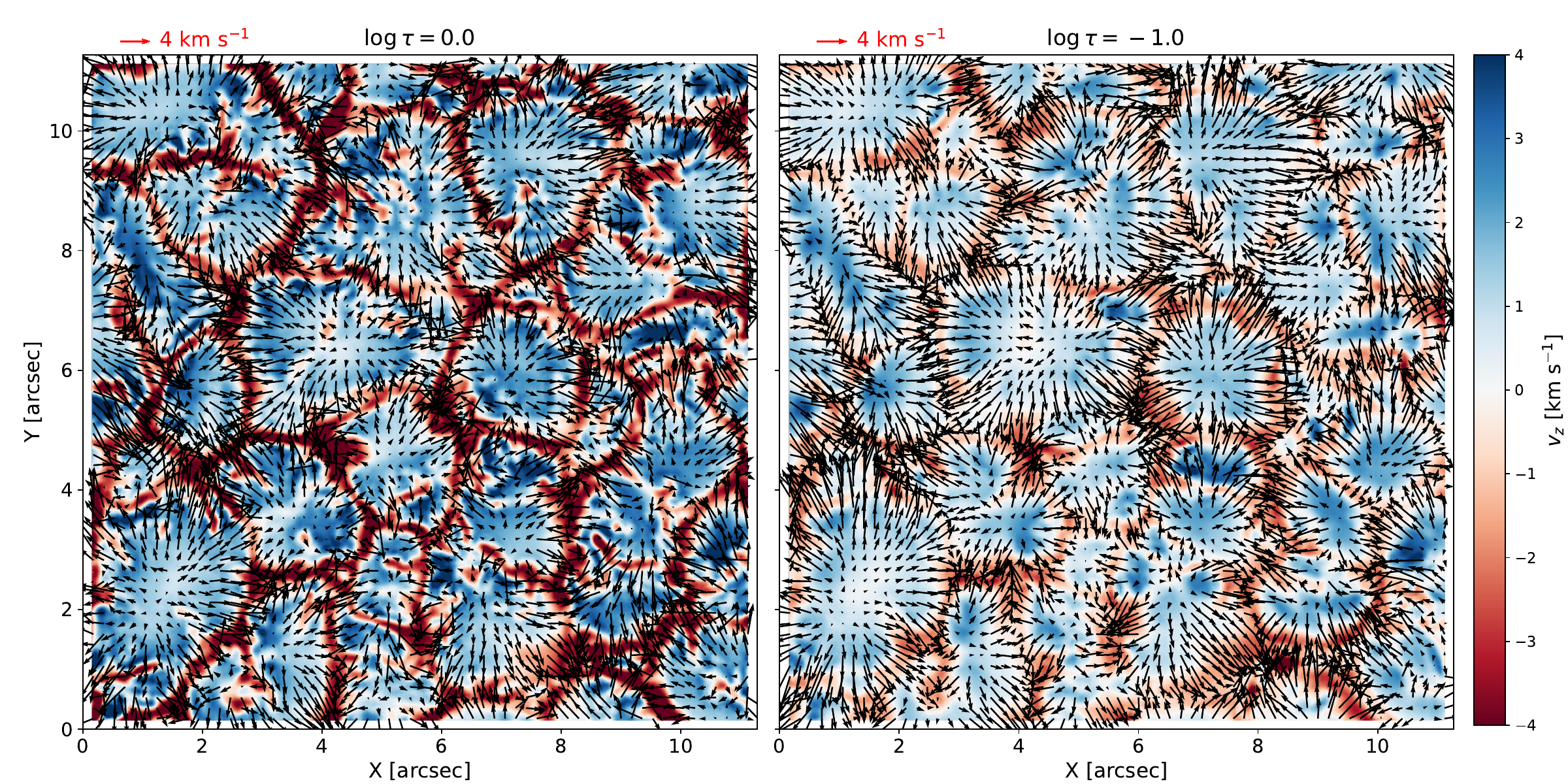}
  \caption{The inferred velocity field at $\log \tau = 0$ (left) and $\log \tau = -1$ (right) with input temporal resolution of $\Delta t = 2$~s. The horizontal arrows indicate the direction and amplitude of horizontal velocity. The vertical velocities are plotted as the background.} 
  \label{fig:dave4vmwdv}
  \vspace{2mm}
\end{figure*}


\subsection{Performance of DAVE4VMwDV at Different Optical Depths}\label{subsec:Performance}

In this section, we investigate the performance of DAVE4VMwDV at different optical depths, similar to the analysis on real data. Here and after we choose to focus on two layers, $\log \tau = 0$ and $\log \tau = -1$, because these two layers roughly bracket the range of formation height for the \ion{Fe}{1} $630$~nm lines. These two layers correspond to the height of approximately $h = 0$~{km} and $h = 135$~{km} above the photosphere according to the widely adopted solar atmospheric model FALC \citep{fal_model}. To extract the maps from the MURaM simulation, we interpolate the magnetic and velocity fields to the $\tau$ space.

To quantify the overall performance, we evaluate the Pearson and Spearman correlation coefficients between all three components of the inferred velocity ($\bm{v}_{\text{DV}}$) and the MURaM ground truth velocity ($\bm{v}_{\text{GT}}$). We further consider three other metrics following the analyses in \cite{schrijver} and \cite{Tremblay}: the spatially averaged relative error
\begin{equation}
  E_{\text{rel}}[\bm{v}_{\text{DV}}, \bm{v}_{\text{GT}}] = \langle \sqrt{\frac{(\bm{v}_{\text{DV}} - \bm{v}_{\text{GT}} )^2}{\bm{v}_{\text{DV}} \cdot \bm{v}_{\text{GT}}}} \rangle,
\end{equation}
the vector correlation coefficient
\begin{equation}
  C[\bm{v}_{\text{DV}}, \bm{v}_{\text{GT}}] =  \frac{\langle \bm{v}_{\text{DV}} \cdot \bm{v}_{\text{GT}}\rangle}{ \langle \lVert \bm{v}_{\text{DV}} \rVert _2 \rangle \langle \lVert\bm{v}_{\text{GT}} \rVert _2\rangle },
\end{equation}
and the cosine similarity index that measures the average cosine of the angle between ground truth and inferred velocity
\begin{equation}
  A[\bm{v}_{\text{DV}}, \bm{v}_{\text{GT}}] = \langle \frac{\bm{v}_{\text{DV}} \cdot \bm{v}_{\text{GT}}}{\lVert \bm{v}_{\text{DV}} \rVert _2 \lVert\bm{v}_{\text{GT}} \rVert _2} \rangle .
\end{equation}
The perfect inferred velocities should have $E_{\text{rel}} = 0$, $C = 1.0$, and $A = 1.0$.

In this test, we use the vector magnetic fields and Doppler velocity at $t = 2$~s as input for each optical depth. The time derivative of $\partial B_z/ \partial t$ at $t = 2$~s is calculated from $B_z$ at $t = 0$~s and $t = 4$~s.  The uncertainties for $\partial B_z/ \partial t$ is set as the difference between $\partial B_z/ \partial t$ and the ideal induction equation (non-ideal effects owing to, e.g., intrinsic numerical errors of simulation, spatial/spectral binning), and the uncertainty for $v_l$ is taken to be a typical observational value $200$~{m~s$^{-1}$}. For the DAVE4VMwDV free parameters, we set the window size to be $w = 15$, and the degree of Legendre expansion in the horizontal and vertical directions are $d=5$ and $d_r=7$ for $\log \tau = 0.0$ and $d=3$ and $d_r=5$ for $\log \tau = -1.0$, respectively. The choice of free parameters is discussed in Appendix \ref{app:opt_MHD}.

The maps of the inferred velocity field are shown in Figure~\ref{fig:dave4vmwdv}. Both layers show the expected convective flow pattern: diverging in the granular cell centers and converging in the intergranular lanes. The inferred velocities are generally smoother than ground-truth velocities (cf. Figure~\ref{fig:MURaM}). Some fine structures, in particular narrow downflow lanes extending from the intergranular lane to the granular cell center, are not well reproduced.

\begin{figure*}[t!]
  \centering
  \includegraphics[width=0.95\textwidth]{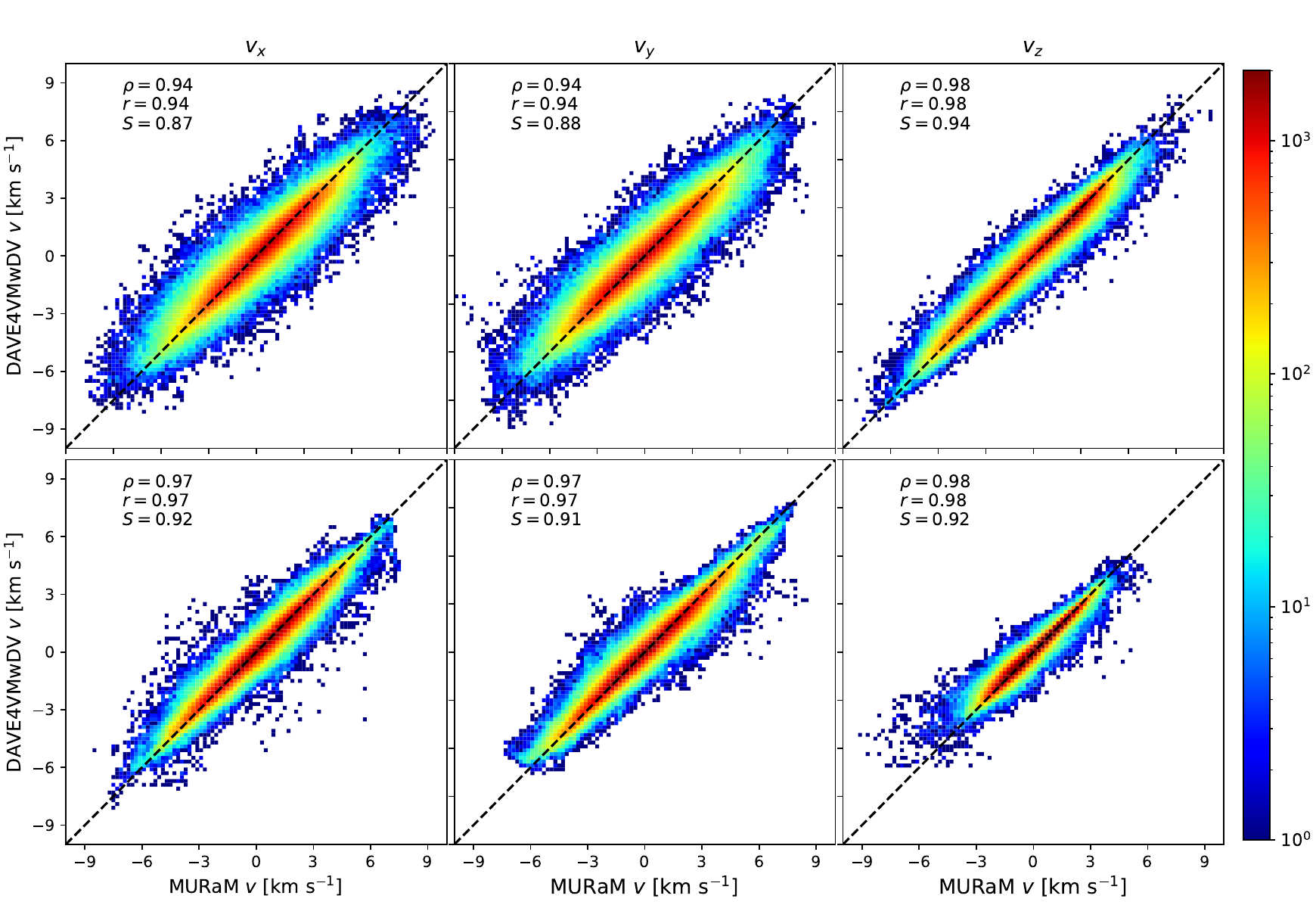}
  \caption{2D histograms of the inferred velocity field and the reference velocity field at $\log \tau = 0$ (top) and $\log \tau = -1$ (bottom). From left to right, we show the histograms for $v_x$, $v_y$, and $v_z$, respectively. The Spearman coefficient ($\rho$), Pearson coefficient ($r$), and slope ($S$) are also shown on the plots.}
  \label{fig:vel_hist}
  \vspace{2mm}
\end{figure*}

The scatter plots between inferred and the ground-truth velocities at two optical depths are shown in Figure~\ref{fig:vel_hist}. The inferred velocity at $\log \tau = 0$ has $E_{\text{rel}} = 0.28$, $C = 0.95$, and $A = 0.95$, while the inferred velocity at $\log \tau = -1$ has $E_{\text{rel}} = 0.23$, $C = 0.97$, and $A = 0.96$. It appears that DAVE4VMwDV has better performance on the higher layer, which has less vigorous convection and thus less complex spatial structure.

Despite all three components of the inferred velocities having Pearson/Spearman coefficients greater than $0.9$ at both layers, the slopes between the inferred velocities and ground truth velocities are smaller than $1$, suggesting an overall underestimate of the flow fields. The mean magnitude of inferred velocity $v=\sqrt{v_x^2+v_y^2+v_z^2}$ at $\log \tau = 0$ and $\log \tau = -1$ are $3.4$~{km~s$^{-1}$} and $2.9$~{km~s$^{-1}$}, respectively, compared to $3.8$~{km~s$^{-1}$} and $3.1$~{km~s$^{-1}$} from the ground-truth.

\begin{figure*}[t!]
  \centering
  \includegraphics[width=0.95\textwidth]{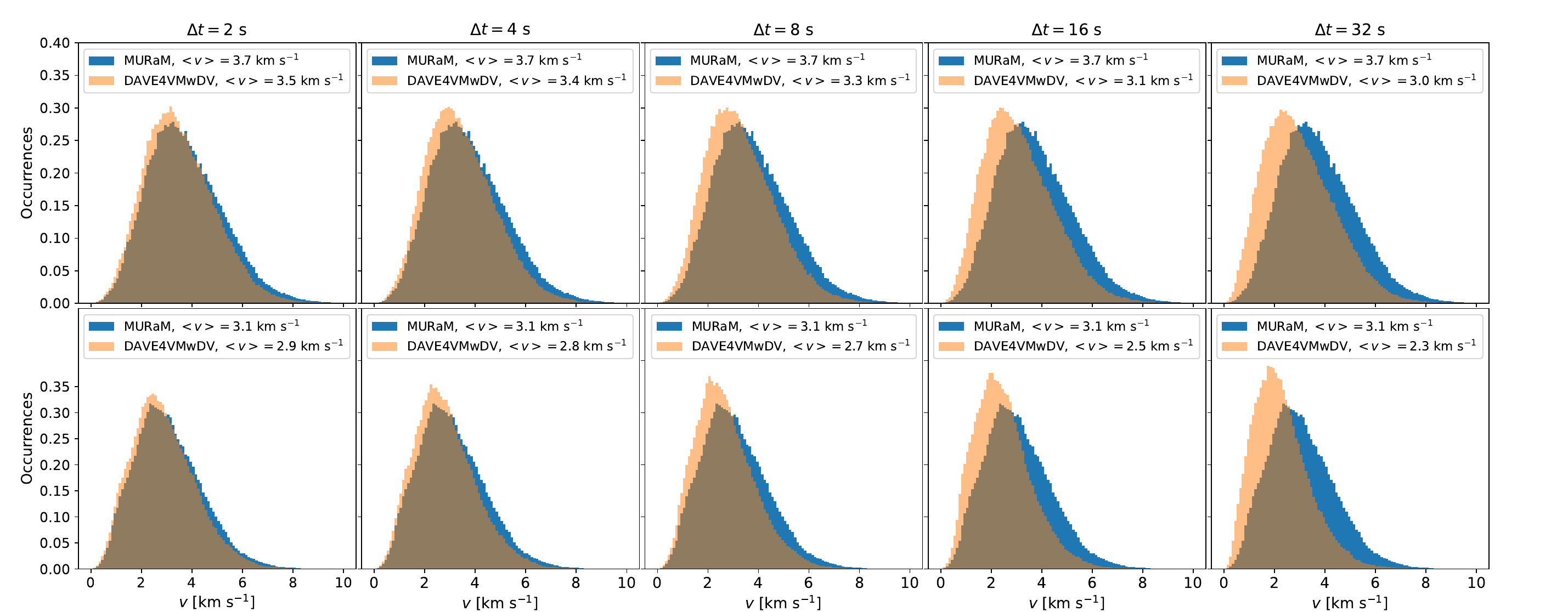}
  \caption{Histograms of the magnitude of the ground-truth velocity (blue) and inferred velocity (orange) from magnetograms with different temporal resolution. Top: histograms for $\log \tau = 0$. Bottom: histograms for $\log \tau = -1$. From left to right, we show the results with $\Delta t = [2, 4, 8, 16, 32]$ s, respectively.}
  \label{fig:vel_histograms}
  \vspace{2mm}
\end{figure*}

\subsection{The Effect of Temporal Resolution on Velocity Estimate}

We apply DAVE4VMwDV on MURaM data with five different cadences to investigate the effect of temporal resolution. The input parameters are the magnetograms and Dopplergram at $t_0 = 32$~s at $\log \tau = 0$ and $\log \tau = -1$. The time derivatives of vertical magnetic field are calculated from vertical magnetic fields at $t_1 = [30, 28, 24, 16, 0]$ s and $t_2 = [34, 36, 40, 48, 64]$~s, so that the nominal cadences are $\Delta t = [2, 4, 8, 16, 32]$~s.

The histograms of the velocity magnitudes inferred with various cadences are shown in Figure \ref{fig:vel_histograms}. The overall magnitude is always smaller than the ground truth, even in the case of the highest cadence. As the cadence decreases (greater $\Delta t$), the distribution of inferred velocity magnitude becomes narrower, and deviates more from the ground truth for both layers. In the case with $\Delta t = 32$ s, the average velocity magnitude decreases to $3.0$ km s$^{-1}$ and $2.3$ km s$^{-1}$ at $\log \tau = 0$ and $\log \tau = -1$, compared to ground truth values of $3.7$ km s$^{-1}$ and $3.1$ km s$^{-1}$. 

\begin{figure}[t!]
  \centering
  \includegraphics[width=0.45\textwidth]{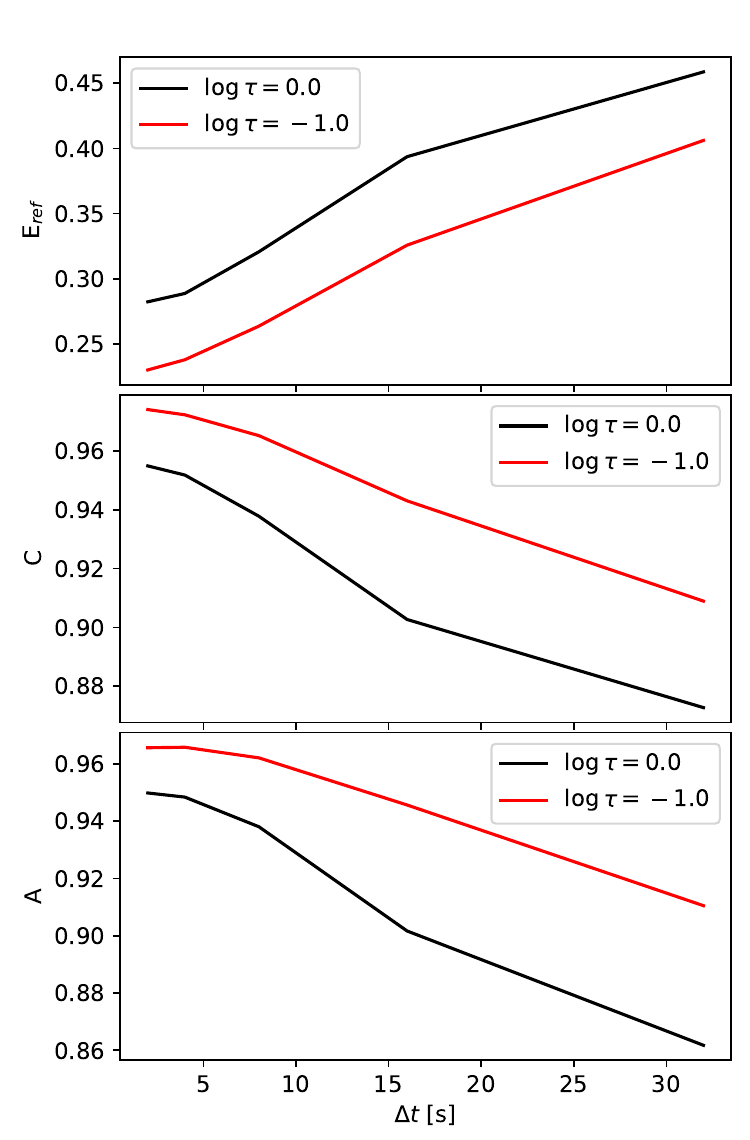}
  \caption{Variation of metrics $E_{\text{rel}}$, $C$ and $A$ (from top to bottom) with respect to cadence $\Delta t$. The black lines represent to the metrics at $\log \tau = 0.0$, and the red lines represent to the metrics at $\log \tau = -1.0$. }
  \label{fig:metrics}
  \vspace{2mm}
\end{figure}

Figure \ref{fig:metrics} shows for performance metrics $E_{\text{rel}}$, $C$, and $A$ for velocity at $\log \tau = 0$ (black) and $\log \tau = -1$ (red). The model performance is consistently better at the higher layer as aforementioned. As $\Delta t$ increases, all three metrics become worse. We posit that the worsening performance after $\Delta t \ge 4$~s may be attributed to the violation of the Courant-Friedrichs-Lewy (CFL) condition. The average horizontal velocities are $v_h = 2.9$ km s$^{-1}$ and $2.8$ km s$^{-1}$ at $\log \tau = 0$ and $-1$, respectively. For the MURaM horizontal grid size of $16$~km, a cadence of $\Delta t = 2.5$~s or better is required.

\begin{figure*}[t!]
  \centering
  \includegraphics[width=0.98\textwidth]{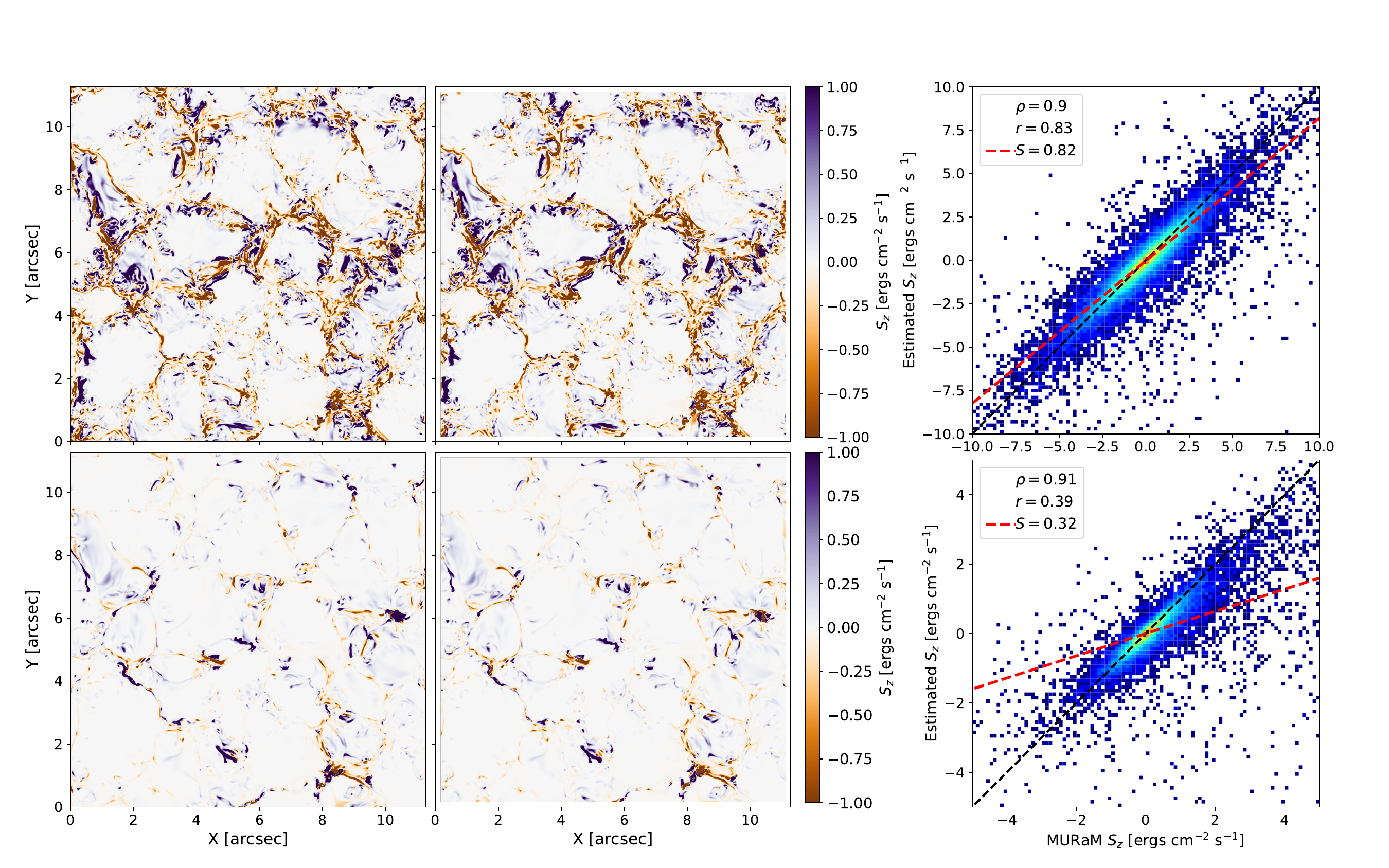}
  \caption{Comparison between Poynting flux from MHD simulation (left) and DAVE4VMwDV (middle) at $\log \tau = 0$ (top) and $\log \tau = -1$ (bottom). The right column shows the 2D histogram between reference and estimate Poynting flux. The Spearman coefficient ($\rho$), Pearson coefficient ($r$), and linear fitting (red dashed line) between ground truth and estimated Poynting flux are shown on the 2D histogram.}
  \label{fig:poy_MHD}
  \vspace{2mm}
\end{figure*}


\begin{deluxetable*}{ccq{3.1}q{3.1}q{3.1}q{3.1}q{3.1}}[t!]
  \tablecaption{Summary of estimated Poynting flux}\label{tab:Poynting}
  \tablecolumns{7}
  \tablewidth{0pt}
  \tablehead{
  \colhead{Optical Depth} & \colhead{Source} & \colhead{$\left<S_z\right>$} & \colhead{$\left<\lvert S_z \rvert\right>$} & \colhead{$\left<S_z^\text{em}\right>$} & \colhead{$\left<S_z^\text{sh}\right>$} & \colhead{$\left<\lvert S_z^\text{sh}\rvert\right>$} 
  }
  \startdata
  $\log \tau = 0.0$ & MURaM, Ground Truth  & 3.7 & 278.9 & -68.4 & 72.1 & 201.1  \\ 
    &  MURaM, DAVE4VMwdV    & -16.2 & 269.4 & -68.1 & 51.9 & 175.5 \\ 
    & Emulation, DAVE4VMwdV & -10.5 & 150.5 & -23.0 & 12.5 & 95.2\\ 
  $\log \tau = -1.0$ & MURaM, Ground Truth & 30.2 & 75.7 & -27.8 & 58.0 & 75.3 \\ 
    & MURaM, DAVE4VMwdV     & 7.4 & 64.0 & -24.7 & 32.2 & 53.6 \\ 
    & Emulation, DAVE4VMwdV & -8.6 & 36.2 & -13.7 & 5.1 & 22.7 \\ 
  \enddata
  \tablecomments{The unit for all values is $10^6$~ergs~cm$^{-2}$~s$^{-1}$. The quantities from left to right are net Poynting flux $\left<S_z\right>$, averaged unsigned Poynting flux $\left<\lvert S_z \rvert\right>$, averaged emergence term of Poynting flux $\left<S_z^\text{em}\right>$, averaged shearing term of Poynting flux $\left<S_z^\text{sh}\right>$, and averaged unsigned shearing term of Poynting flux $\left<\lvert S_z^\text{sh}\rvert\right>$. }
  \end{deluxetable*}
  

\subsection{Estimation of Poynting Flux}\label{subsec:Poynting}

Based on the velocity inferred by DAVE4VMwDV, we may estimate the Poynting flux through the surface with the Equation~\ref{equ:Poynting}. The estimated Poynting fluxes are listed in Table \ref{tab:Poynting}. The ground-truth and the estimated Poynting flux maps are shown in the left and middle columns of Figure~\ref{fig:poy_MHD}, respectively. The inference results qualitatively recovers the pattern of ground truth, with large $S_z$ values concentrated near the intergranular lanes.

While Poynting flux of both signs exists near the intergranular lanes, there are overall more negative occurrences. In contrast, the granular cell centers have relatively uniform but weak, positive Poynting flux. From $\log \tau = 0$ to $-1$, the fractional area occupied by large Poynting flux decreases with height, and the pattern becomes less structured, reminiscent of the flow maps.


The average \emph{unsigned} Poynting flux is greater at the lower layer: the value for $\log \tau = 0$ is $2.8 \times 10^8$ ergs cm$^{-2}$ s$^{-1}$, while for $\log \tau = -1$ it is $0.8 \times 10^8$ ergs cm$^{-2}$ s$^{-1}$. Quite interestingly, the \emph{net} Poynting flux is greater at the higher layer: the value for $\log \tau = 0$ is $3.7 \times 10^6$ ergs cm$^{-2}$ s$^{-1}$, while for $\log \tau = -1$ is $3.0 \times 10^7$ ergs cm$^{-2}$ s$^{-1}$, almost is an order of magnitude larger. This is because the convection is stronger in the lower layer so that magnitude of vertical velocity is larger bringing more downward Poynting flux, which cancels out the upward Poynting flux.  

We compare the inferred values to the ground-truth. The estimated unsigned Poynting flux can recover $96.4\%$ and $84.2\% $ of the ground truth unsigned Poynting flux at $\log \tau = 0$ and $\log \tau = -1$, respectively. However, the estimated net Poynting flux at these two layers are $-1.6 \times 10^7$ ergs cm$^{-2}$ s$^{-1}$ and $7.4 \times 10^6$ ergs cm$^{-2}$ s$^{-1}$, respectively. They are quite far from the ground truth in magnitude, and have the wrong sign for $\log \tau = 0.0$. We will discuss below the cause of the mismatch.

The last column of Figure~\ref{fig:poy_MHD} shows the two-dimensional histograms of Poynting flux between the ground truth and our estimate. While there is overall reasonable correlation between the two variables, significant scatter is apparent for both heights. To investigate the origin of the scatter, we consider the Poynting flux in regions with $\lvert B \rvert \ge 3\sigma_B$ and regions with $\lvert B \rvert < 3\sigma_B$ separately, where $\sigma_B$ here stands for the standard deviation in the FOV. The stronger-field region represents $34.4\%$ and $50.5\%$ of the total unsigned Poynting flux at $\log \tau = 0$ and $\log \tau = -1$, respectively. In these weak-field regions, the Pearson correlation coefficients between the inferred and the ground-truth velocities are $0.91$ and $0.85$ for the lower and higher layers, respectively. Counterintuitively, the correlation coefficients in stronger-field regions are only $0.78$ and $0.36$, suggesting that the scatter mostly results from the stronger-field pixels.

We note that a large fraction of pixels are below the one-to-one line. In weak-field regions, the net Poynting fluxes are $-4.3\times 10^6$ ergs cm$^{-2}$ s$^{-1}$ and $6.0\times 10^6$ ergs cm$^{-2}$ s$^{-1}$ at these two layers, which are smaller compared to the ground truth value of $8.6\times 10^5$ ergs cm$^{-2}$ s$^{-1}$ and $9.1\times 10^6$ ergs cm$^{-2}$ s$^{-1}$. The Poynting flux in stronger-field regions is significantly greater and also underestimated, which accounts for the majority of the discrepancy. The net Poynting fluxes in the stronger-field region are $-3.5\times 10^8$ ergs cm$^{-2}$ s$^{-1}$ and $7.3\times 10^7$ ergs cm$^{-2}$ s$^{-1}$ at these two layers, compared to the ground truth value of $1.3\times 10^8$ ergs cm$^{-2}$ s$^{-1}$ and $9.8\times 10^8$ ergs cm$^{-2}$ s$^{-1}$.

We further decompose the Poynting flux into an emerging term and a shearing term. The values of these two terms are also listed in Table \ref{tab:Poynting}. The ground-truth, net emerging term and net shearing term have different signs at both layers. These two terms almost cancel out at $\log \tau = 0.0$. In contrast, the shearing term dominates over the emergence term at $\log \tau = -1.0$: the absolute value is about twice as large. We find that the estimated net emerging Poynting flux can reproduce $99.6\%$ and $88.8\%$ of the ground truth at $\log \tau = 0$ and $\log \tau = -1$, respectively. But the estimate can only reproduce $72.0\%$ and $55.5\%$ of the net shearing term. The underestimation of the shearing term accounts for the majority of the mismatch of Poynting flux.


\section{Estimated Energy Transport from Emulated Observation} \label{sec:synthesis}

Above, we focused on the performance of {\dfvmwdv} on MHD simulation data. In this section, we examine the capability of estimating energy transport in the emulated DL-NIRSP observation. As we shall see, most of the conclusions in the previous section qualitatively hold.

\subsection{Stokes Synthesis and MHD Variables Inference}\label{sec:inversion}

\begin{figure}[t!]
  \centering
  \includegraphics[width=0.48\textwidth]{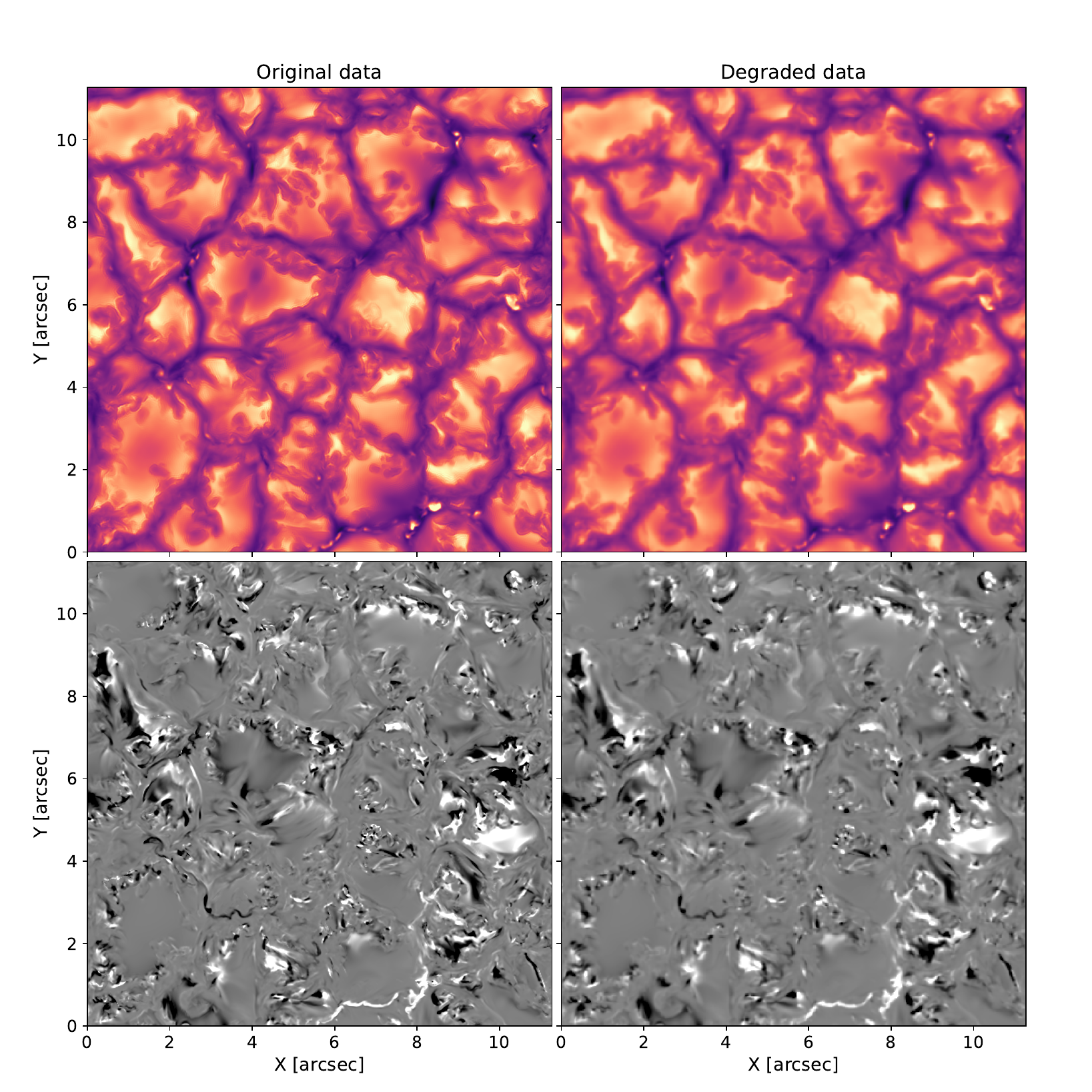}
  \caption{Original (left) and degraded (right) synthetic Stokes profiles. Upper panels show continuum intensity $I/I_c$, and lower panels show the circular polarization $V/I_c$ in the wing of \ion{Fe}{1} 630.15~nm line.} 
  \label{fig:degraded}
  \vspace{2mm}
\end{figure}

To emulate the DL-NIRSP observation, we first synthesize the four Stokes profiles from the MURaM simulation with SIR and then degrade the profiles with the theoretical point spread function of DL-NIRSP in the High-Res mode and a two-pixel binning. Figure \ref{fig:degraded} shows an example of the synthesized data in the original and degraded resolution. After degradation, the contrast in the continuum image decreases by $9\%$. In the Stokes $V$ map, the original mixed polarity features in the intergranular lane become weaker due to cancellation; lanes with stronger signals become wider.

\begin{figure}[t!]
  \centering
  \includegraphics[width=0.45\textwidth]{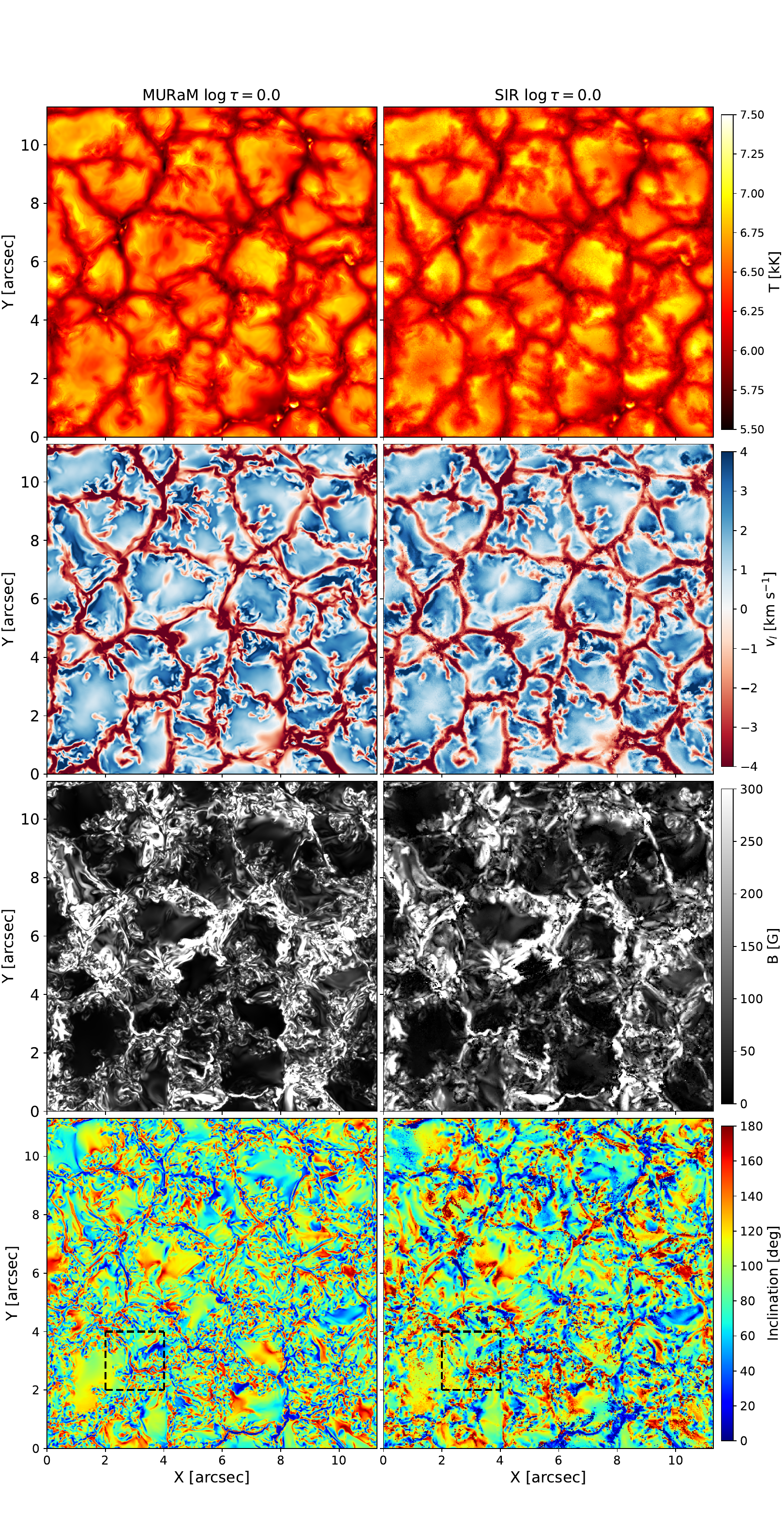}
  \caption{Comparison between the ground-truth atmosphere (left) and that inferred with SIR (right) at $\log \tau = 0$. From top to bottom, we show temperature, LOS velocity, magnetic field strength, and inclination, respectively. An example of a mixed-polarity region is marked with black dashed box.}
  \label{fig:SIR_0.0}
  \vspace{2mm}
\end{figure}

\begin{figure}[t!]
  \centering
  \includegraphics[width=0.45\textwidth]{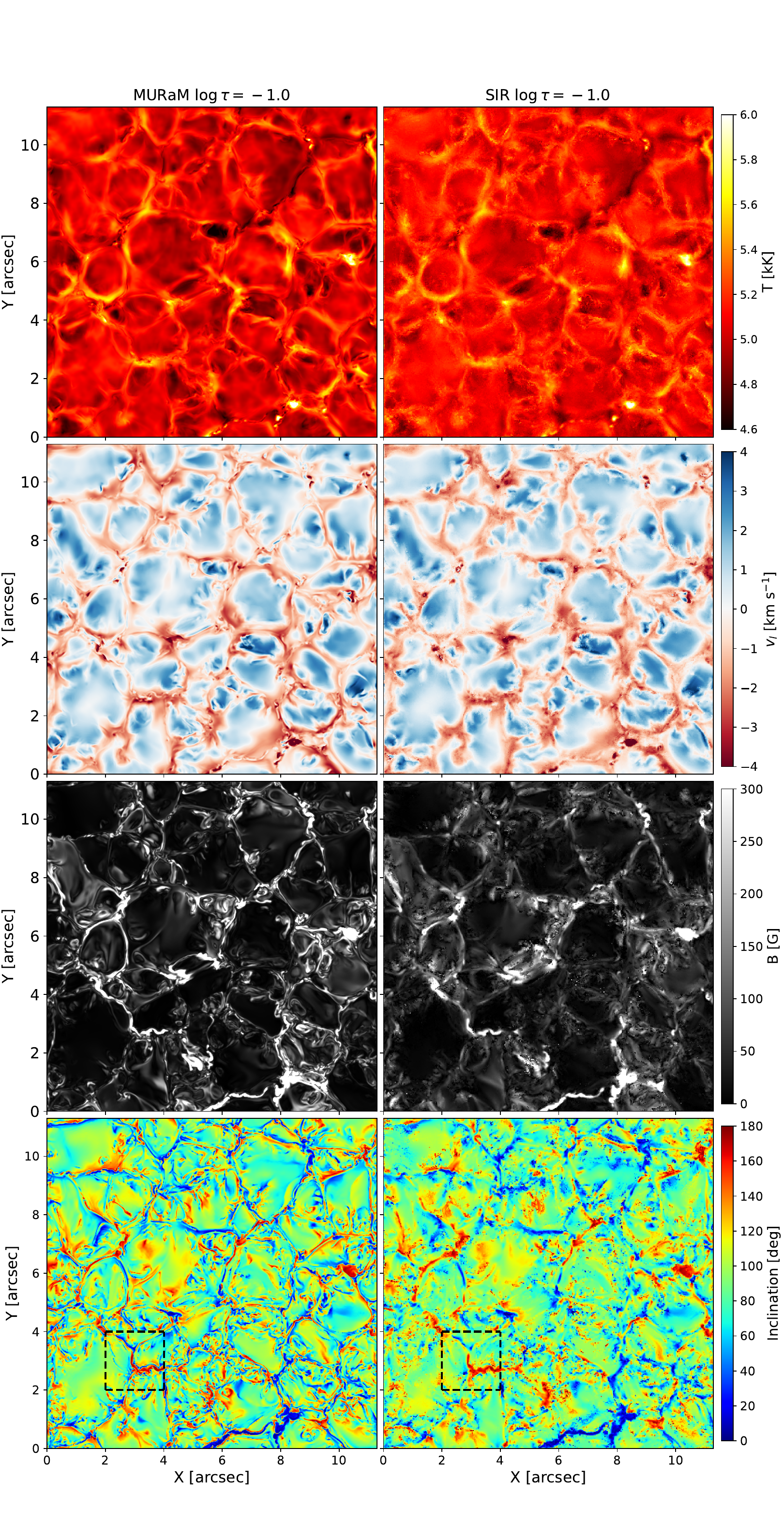}
  \caption{Same as Figure \ref{fig:SIR_0.0} but for $\log\tau = -1$.} 
  \label{fig:SIR_-1.0}
  \vspace{2mm}
\end{figure}

The comparison between input MURaM parameters and inferred parameters with SIR at $\log \tau = 0$ and $\log \tau = -1$ is shown in Figure \ref{fig:SIR_0.0} and Figure \ref{fig:SIR_-1.0}. 

At $\log \tau = 0$, the inferred temperature map and LOS velocity map have no appreciable difference from the ground truth MURaM map. They both show typical granular structure: high temperature and upflows in granular cell centers, with low temperature and downflows in intergranular lanes. The continuum bright points in intergranular lanes are also captured. The inferred magnetic field strength shows a network-like pattern similar to the ground truth, though fine-scale structures are diminished. We will discuss below the response of inferred magnetic field to spatial scales. The inferred inclination captures the large-scale pattern with more vertical magnetic fields filling the intergranular lanes. The small-scale mixed polarity pattern, for example, the region centered at $(X, Y) = (3\arcsec, 3\arcsec)$ in the black dashed box, is largely missing.

At $\log \tau = -1$, the inferred temperature map appears to be overall brighter than the ground truth map. The average inferred temperature is $5122$~K, about $58$ K higher than the ground truth, which is similar to previous work \citep{quiteronoda2023}. The difference between the inferred and ground-truth LOS velocity map, on the other hand, appears to be minimal. The inferred magnetic field strength shows a similar pattern to ground truth, albeit somewhat blurry. The average field strength in the strong field region ($B > 100$ G) is $195$ G, compared to the ground truth value of $230$ G. The inferred inclination fails to capture the small-scale mixed polarities, similar to the lower layer.


\begin{figure}[t!]
  \centering
  \includegraphics[width=0.45\textwidth]{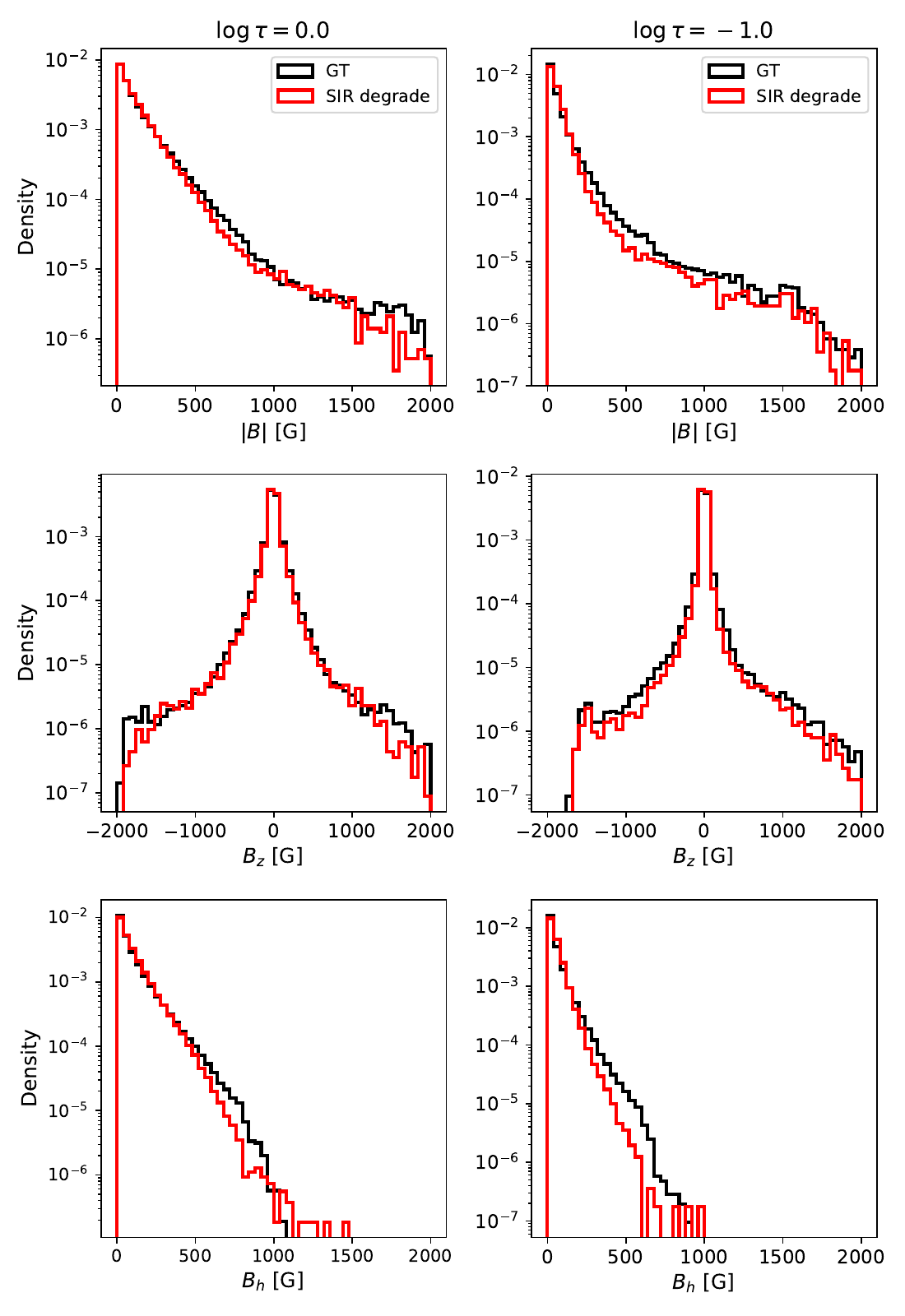}
  \caption{Comparison of the distributions of the ground-truth (black) and inverted magnetic field (red). Left: Histograms of quantities at $\log \tau = 0$. Right: Histograms of quantities at $\log \tau = -1$. The figure shows the distribution of magnetic field strength $\lvert B \rvert$, vertical magnetic field $B_z$, and horizontal magnetic field $B_h$. The inversion results have lower resolution and thus fewer pixels, and the histograms are all normalized for comparison.} 
  \label{fig:histogram}
  \vspace{2mm}
\end{figure}

We compare the distributions of the total ($|B|$), vertical ($B_z$), and horizontal ($B_h$) magnetic fields in Figure \ref{fig:histogram}. The deviation between the ground truth and the inversion results appears to be more pronounced for strong field pixels and for the higher layer at $\log \tau=-1$. At $\log \tau = 0$, the largest deviation appears to be an underestimate of $B_h$ for values above $500$~G. For $\log \tau = -1$, the inferred fields are generally significantly weaker than the ground truth.



\begin{figure*}[t!]
  \centering
  \includegraphics[width=0.95\textwidth]{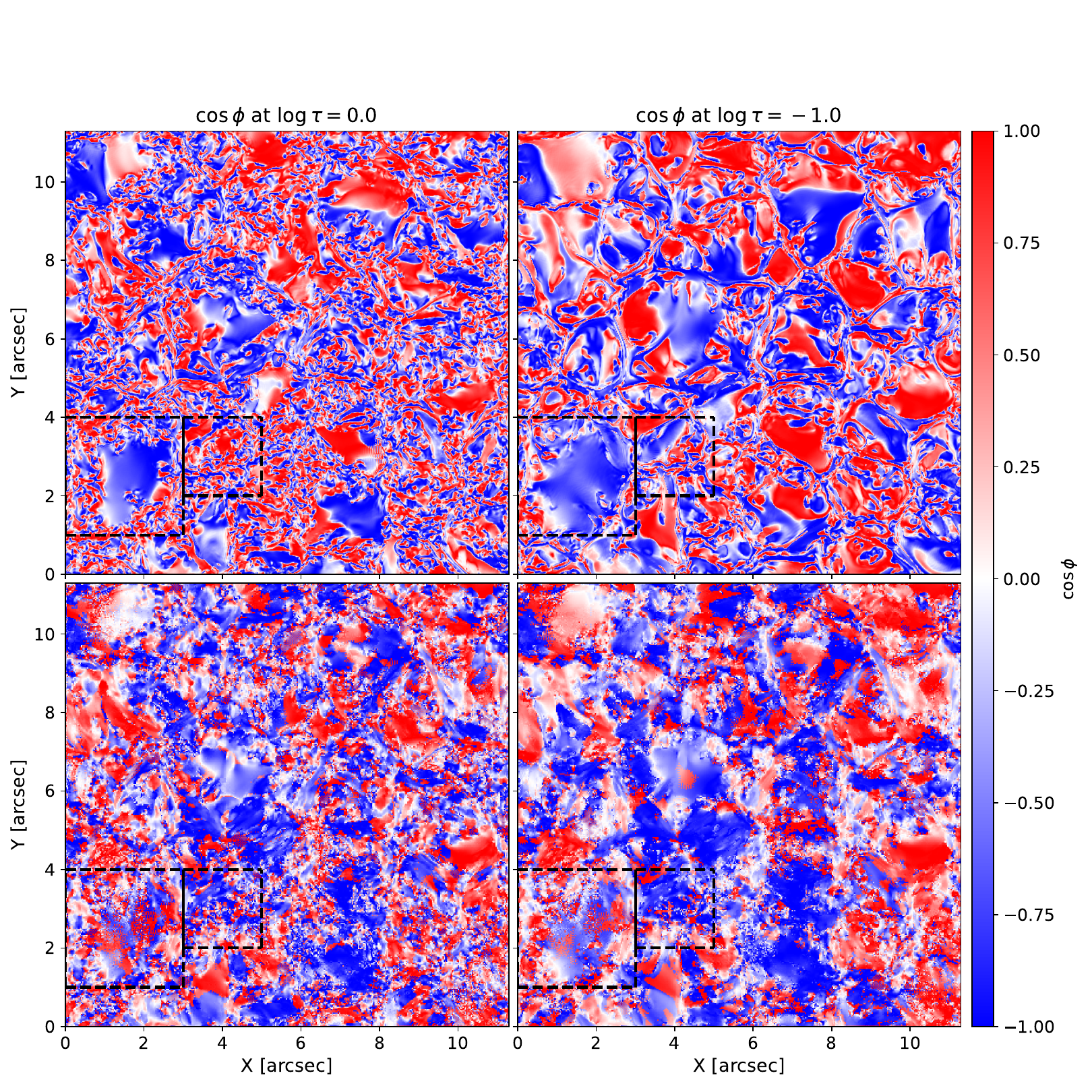}
  \caption{Comparison of the cosine of azimuth between ground truth (upper row) and the one inferred with SIR after disambiguation (bottom row). From left to right we show the comparison at $\log \tau = 0$ and $\log \tau = -1$. Two dashed black boxes mark regions discussed Section \ref{sec:inversion}} 
  \label{fig:azimuth}
  \vspace{2mm}
\end{figure*}

Inaccuracies may also originate from the azimuthal disambiguation step. Figure~\ref{fig:azimuth} illustrates the cosine of azimuth ($\cos \phi$) from the MURaM model (left), versus the inferred results after the application of the ME0 algorithm (right). As evidenced by the large mismatch, the performance of the ME0 algorithm is not satisfactory. In the granular lanes, for example, the region centered at $(X, Y) = (4\arcsec, 3\arcsec)$ in the black box, fine-scale structures where the sign of $\cos \phi$ changes within several pixels are not reproduced. The results in granular cell centers, for example, the granular cell centered at $(X, Y) = (1.5\arcsec, 2.5\arcsec)$ and marked by a white cross, where $\phi$ is expected to vary smoothly, are somewhat better, though patches with completely opposite sign also exist. The solution of the ME0 disambiguation algorithm is known to be unsatisfactory for the quiet Sun: similar conclusions can be found in recent work \citep[e.g.,][]{Tilipman}.


\subsection{Poynting Flux from Emulated Observation}

We use the inverted magnetic field and Doppler velocity maps to infer the 3D velocity field with DAVE4VMwDV. As shown in Section \ref{sec:DAVE4VMwDV}, higher cadence is preferred. To access the best performance of existing schemes, we will use inferred vector magnetic fields and Doppler velocity at $t = 2$~s as input for each optical depth. The time derivative of $\partial B_z/ \partial t$ at $t = 2$~s is calculated from $B_z$ at $t = 0$~s and $t = 4$~s. We set window size $w = 15$, degree of expansion $d = 3$ and $d_r = 5$ for $\log \tau = 0$, and set $w = 15$, $d = 1$ and $d_r = 5$ for $\log \tau = -1$. The choice of parameters is discussed in Appendix~\ref{app:optimization}. We then estimate the Poynting flux with the inferred velocity and Equation~\ref{equ:Poynting}.

\begin{figure*}[t!]
  \centering
  \includegraphics[width=0.98\textwidth]{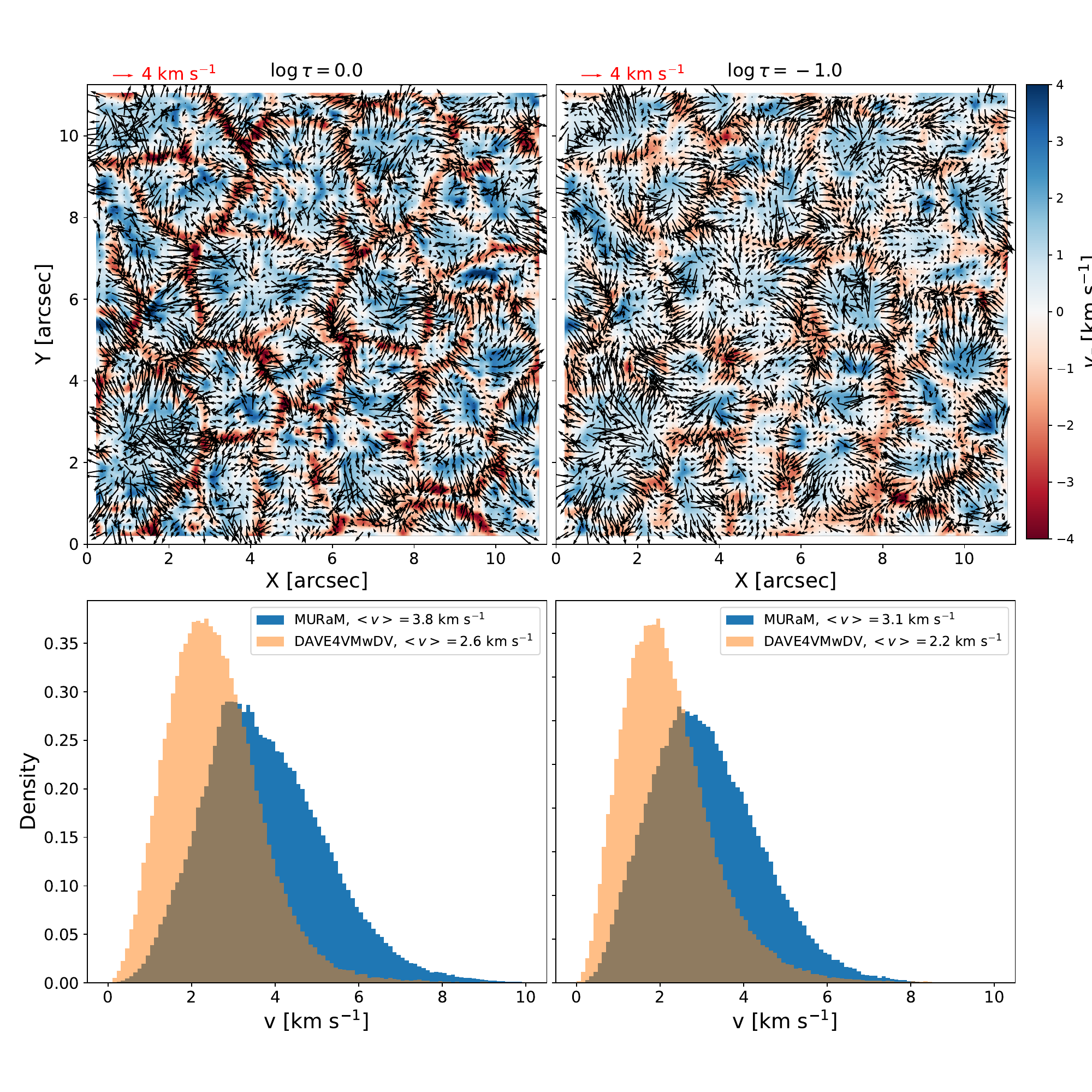}
  \caption{Velocities inferred from inverted magnetograms at $\log \tau = 0.0$ (left) and $\log \tau = -1.0$ (right). Top: The inferred velocity field. The horizontal arrows indicate the direction and amplitude of horizontal velocity. The vertical velocities are plotted as the background. Bottom: Histograms of the magnitude of the ground-truth velocity (blue) and inferred velocity (orange). } 
  \label{fig:vel_SIR}
  \vspace{2mm}
\end{figure*}

The velocity fields inferred from the emulated magnetograms are shown in Figure~\ref{fig:vel_SIR}. Similar to the velocity inferred from the ground-truth magnetograms in the previous section, convective patterns are well recovered for both layers. The flow direction in the granular cell center, on the other hand, appears to be more random and disordered, unlike the consistent diverging flows seen previously. Compared to the distribution of the ground truth velocity, the inferred velocities at both layers have a narrower distribution and smaller magnitude. The average magnitude of the inferred velocity is $2.6$~km s$^{-1}$ and $2.2$~km s$^{-1}$ at $\log \tau = 0.0$ and $\log \tau = -1.0$, respectively, compared to $3.8$~km s$^{-1}$ and $3.1$~km s$^{-1}$ from the ground truth.
 
\begin{figure*}[t!]
  \centering
  \includegraphics[width=0.98\textwidth]{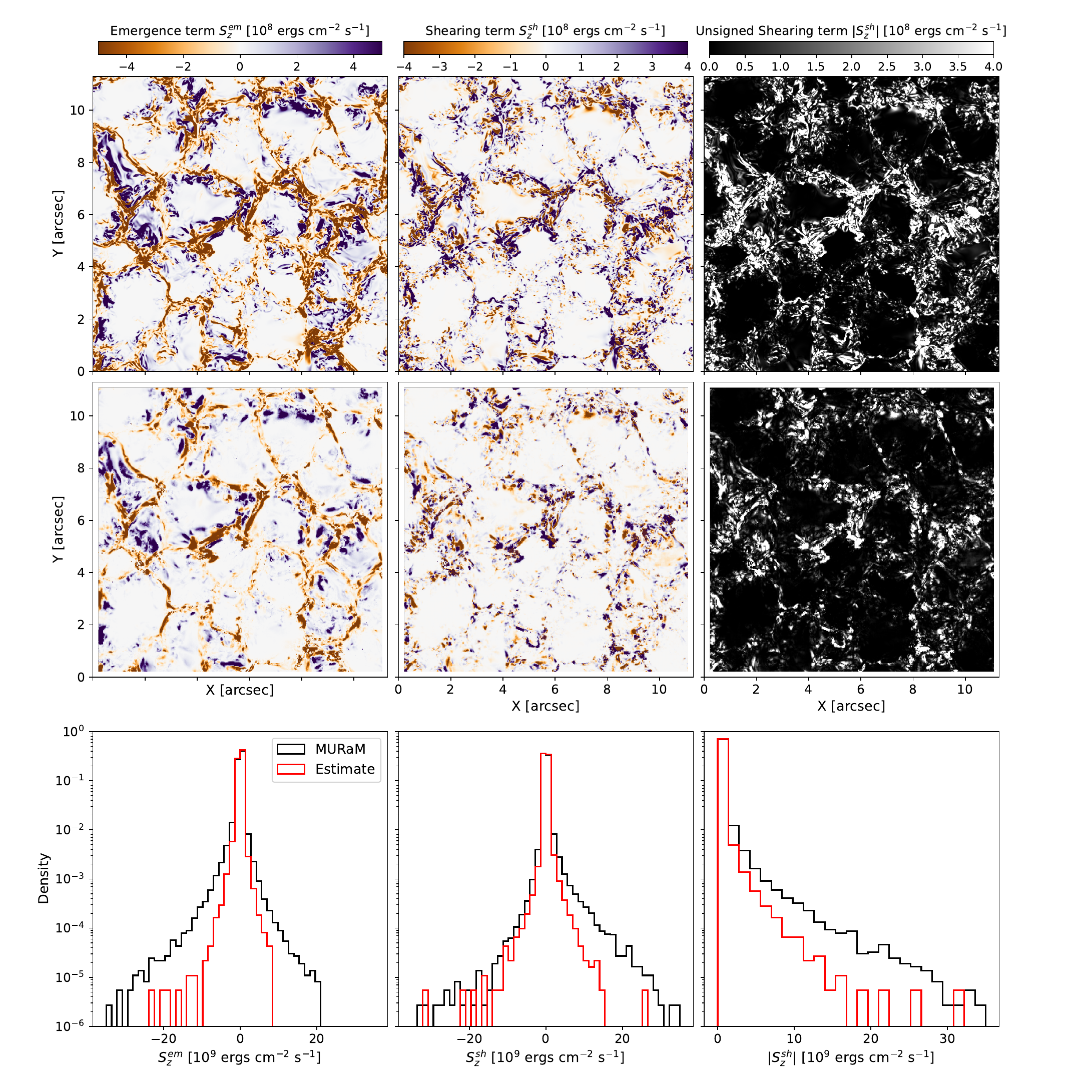}
  \caption{Comparison between MURaM and estimated Poynting flux at $\log \tau = 0$. Top: Ground truth value of the Poynting flux. Middle: Estimated Poynting flux with inferred magnetic and velocity field. Bottom: Histograms of Poynting flux from ground truth (black) and estimate (red). From left to right are the emergence term, shearing term, and unsigned shearing term of Poynting flux. } 
  \label{fig:Poy_0.0}
  \vspace{2mm}
\end{figure*}
\begin{figure*}[t!]
  \centering
  \includegraphics[width=0.98\textwidth]{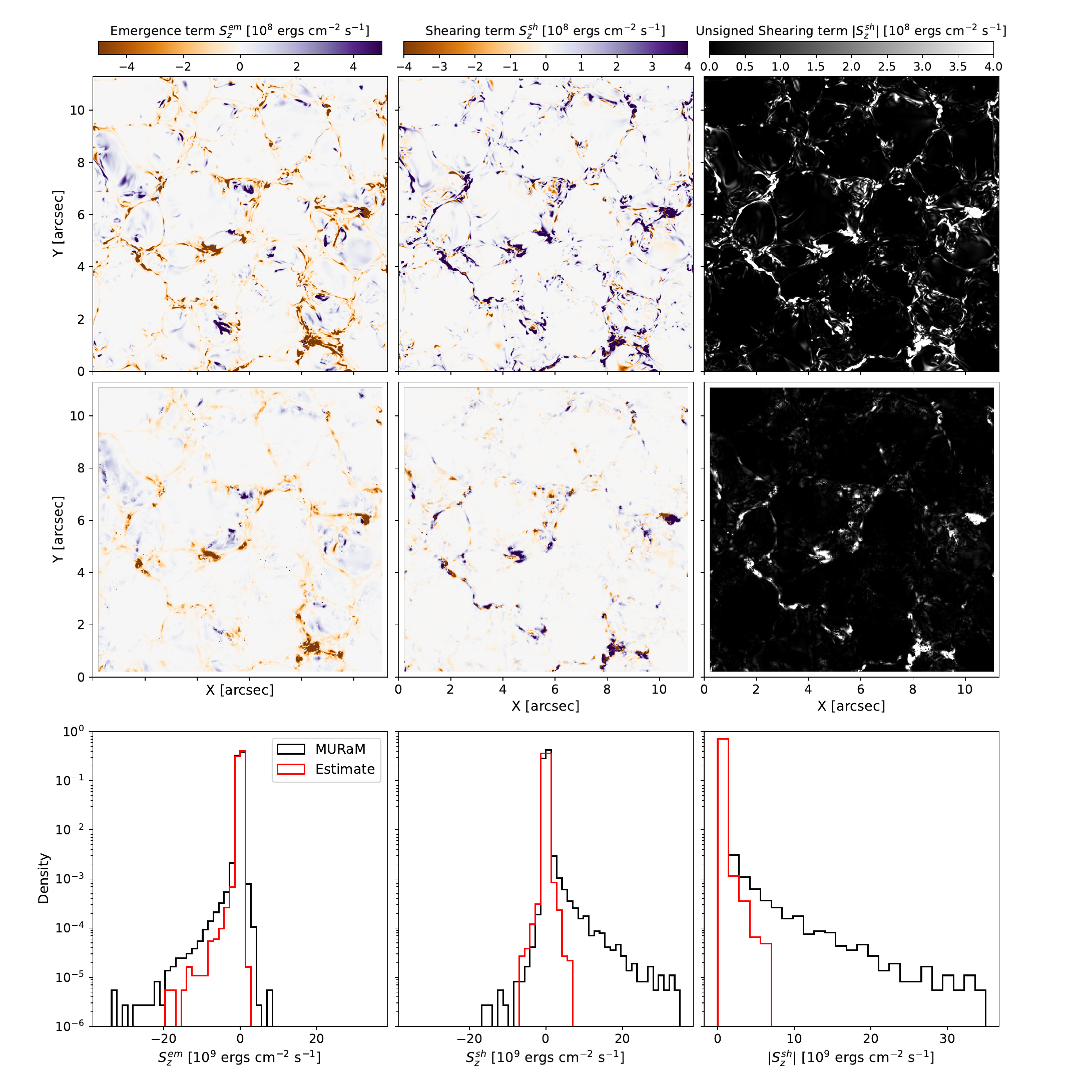}
  \caption{Same as Figure \ref{fig:Poy_0.0}, but for $\log \tau = -1$.}
  \label{fig:Poy_-1.0}
  \vspace{2mm}
\end{figure*}

The estimated Poynting fluxes from emulated magnetograms at the two layers are listed in Table \ref{tab:Poynting}. Below, we examine the estimated emergence term ($S_z^\text{em}$) and shearing term of the Poynting flux ($S_z^\text{sh}$) separately as they are differently affected by the azimuth disambiguation procedure. The emergence term is only related to the square of the horizontal magnetic field ($B_h^2$), so it is not affected by the ambiguity resolution. The shearing term, on the other hand, relates to the vector horizontal magnetic field by $\bm{v}_h \cdot \bm{B}_h$, and can change the sign if the azimuth has the wrong sign. We also examine the unsigned shearing term of Poynting flux ($\lvert S_z^\text{sh} \rvert$) for completeness. The estimate of these three quantities and their comparisons with the ground truth are shown in Figure~\ref{fig:Poy_0.0} and Figure~\ref{fig:Poy_-1.0}.  

At $\log \tau = 0$, the estimated emergence term captures the main features that it is negative in the intergranular lane and positive in the granular cell, though many fine-scale structures disagree. For the estimated shearing term, both the unsigned and signed versions show qualitative agreement with the ground truth: strong values are concentrated along the granular lanes. The absolute values of all three are underestimated overall, as evidenced by the histogram. The signed shearing term, in addition, tends to be more negative than the ground truth. The histograms show an obvious skew towards the negative side. As expected, the estimated Poynting fluxes are always smaller than the ground truth. For the shearing term, the negative fluxes are more consistent with ground truth than positive fluxes, resulting in an overall underestimation. The inferred unsigned Poynting flux is $2.0 \times 10^8$ ergs cm$^{-2}$ s$^{-1}$, which is $53.9\%$ of the ground truth. The emergence term of Poynting flux is $33.6\%$ of the ground truth, while the shearing term of Poynting flux is only $17.3\%$. The unsigned shearing term of Poynting flux is $9.5\times 10^7$ ergs cm$^{-2}$ s$^{-1}$ which is $44.6\%$ of the ground truth. The net Poynting flux is $-1.1 \times 10^7$ ergs cm$^{-2}$ s$^{-1}$, which has a wrong sign and is far from the ground truth value $3.7\times 10^6$ ergs cm$^{-2}$ s$^{-1}$.

Most conclusions for $\log \tau = 0$ hold true for $\log \tau = -1$, while the underestimation becomes more severe. For the shearing term, a large fraction of the strong flux pixels is not recovered. The predominant positive shearing term is also missing. The distribution of $S_z^\textit{sh}$ is much narrower around $0$ compared to the ground truth, with too few positive and a small excess of negative pixels.



\section{Discussion}\label{sec:discussion}

In Section~\ref{sec:DAVE4VMwDV}, we show that DAVE4VMwDV has reasonable performance when applied to the high-resolution simulation data directly. However, the Poynting flux is significantly underestimated. As expected, the case is more severe for the emulated observation case, and we find the main culprit to be the shearing term. We investigate below the quality of the inferred magnetic and velocity fields in an attempt to address the ``missing Poynting flux.''

\subsection{Quality of Inferred Velocity}

DAVE4VMwDV infers the velocity field from magnetograms by minimizing the residual of the ideal induction equation. In our work, we apply DAVE4VMwDV on magnetograms, i.e., magnetic field maps at the same optical depth where the spatial derivatives are poorly defined. Thus, the ideal induction equation does not strictly hold, particularly at small spatial scales where the $\tau$ surface is corrugated. Below, we assess how well the ideal assumption is satisfied and how it affects the inferred velocity at spatial scales.

To this end, we evaluate the power spectral density (PSD) of the inferred variables. For a square FOV in Cartesian coordinates (i.e.\, $n_x = n_y = N$ and $\Delta x = \Delta y$), the power spectral density $PSD(k)$ as a function of the wavenumber is computed as:
\begin{equation}\label{equ:psd}
  PSD(k) = N\Delta x \sum_{k^\prime \in [k,k+dk]} \lvert\tilde{f}(k^\prime)\rvert ^2,
\end{equation}
where $\tilde{f}$ is the two-dimension Fourier transform for an arbitrary spatial function $f$, and $k = \sqrt{k_x^2 + k_y^2}$ is the square root of the wavenumber in the $x$ ($k_x$) and $y$ ($k_y$) direction.

\begin{figure*}[t!]
  \centering
  \includegraphics[width=0.95\textwidth]{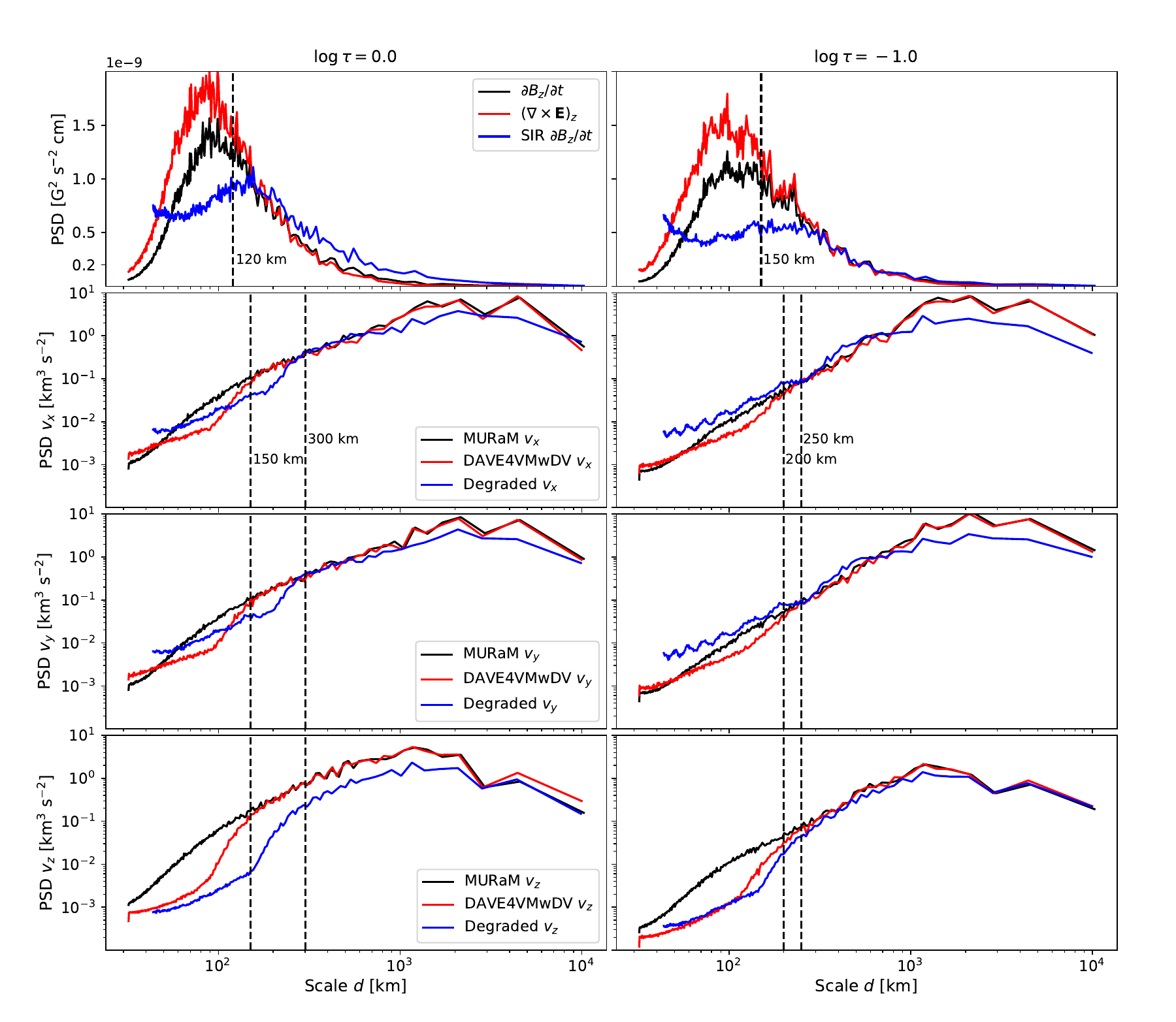}
  \caption{Power spectral density for selective inferred variables as a function of scale $d$ at optical depth $\log \tau = 0$ (left) and $\log \tau = -1$ (right).  From top to bottom: power spectra density of two terms in the induction equation,  $v_x$, $v_y$, and $v_z$. In the top panel, the black line represents the ground-truth time derivative of the vertical magnetic field, the red line represents the vertical component of the curl of the electric field $\bm{E}=-\bm{v}\times\bm{B}$, and the blue line represents the time derivative of {the inferred} vertical magnetic field. In the second to last rows, the black line represents the ground-truth velocity, the red line represents the velocity inferred directly from the MURaM magnetograms (Section~\ref{sec:DAVE4VMwDV}), and the blue line represents the result from velocity inferred from emulated observation (Section~\ref{sec:synthesis}). The PSDs of variables inferred from degraded synthetic data do not reach the smallest scales in the figure due to their lower resolution. The vertical dotted lines highlight the spatial scale, below which large discrepancies exist between the ground truth and the inference.} 
  \label{fig:PSD}
  \vspace{2mm}
\end{figure*}

The PSDs for the two terms of the induction equation, $\partial B_z / \partial t$ and $\left(\bm{\nabla}\times\bm{E}\right)_z$, and the three components of velocity at $\log \tau = 0$ and $\log \tau = -1$ are shown in Figure \ref{fig:PSD}. The values are normalized by their maxima. As seen in the top column, the two terms of the induction equation from the MHD simulation at the same optical depth are not entirely consistent, especially in the scale $d < 120$~{km} (top row, black and red curves). This is because the maps ($\bm{B}$ and $\bm{v}$) we use are at the same $\tau$ surface, which may correspond to different $z$ at different time. The spatial derivatives made in constant $\tau$ surface may also contribute to the mismatch. This significantly affects the quality of inferred velocity based on the sampled ground-truth magnetograms: large discrepancy of PSD in the small-scale regime ($d < 150$~{km}). The agreement is somewhat worse for $\log\tau=-1$ compared to $\log\tau=0$.

For the performance of DAVE4VMwDV on emulated observations, we only plot the PSD for $\partial B_z / \partial t$ in Figure \ref{fig:PSD}. At $\log \tau = 0$, the PSD deviates significantly from the ground truth even at large spatial scales (top row, blue curves). Quite interestingly, the mismatch does not appear to affect much the accuracy of the inferred velocity field on a large scale. The PSDs of all three components of the DAVE4VMwDV velocity closely follow the ground truth for $d < 300$~{km}. Again, the results are similar but better for $\log \tau = -1$. In particular, the PSD of $\partial B_z / \partial t$ agrees well with the reference at scale $d > 250$~{km}.

To quantify the Poynting flux response with respect to the scale of velocity, we decompose the ground truth velocity fields according to their spatial scales:
\begin{equation}
  \begin{split}
    v_z &= v_z^{h} + v_z^{l}, \\
    \bm{v}_h &= \bm{v}_h^{h} + \bm{v}_h^{l},
    \end{split}
\end{equation}
where the superscripts $h$ and $l$ represent the high-frequency part and the low-frequency part of the quantity. The Poynting flux can then be decomposed in a similar fashion as they are linear with respect to the velocity:
\begin{equation}\label{equ:Decomposition}
  \begin{aligned}
      S_z  &= S_z^{h} + S_z^{l}, \\
      S_z^{l} &= \frac{1}{4\pi}\int_S \bm{B}_h^2 v_z^l dS - \frac{1}{4\pi} \int_S (\bm{B}_h\cdot \bm{v}_h^l)B_z dS, \\
      S_z^{h} &= \frac{1}{4\pi}\int_S \bm{B}_h^2 v_z^h dS - \frac{1}{4\pi} \int_S (\bm{B}_h\cdot \bm{v}_h^h)B_z dS. \\
  \end{aligned}
  \end{equation}

\begin{figure}[t!]
  \centering
  \includegraphics[width=0.48\textwidth]{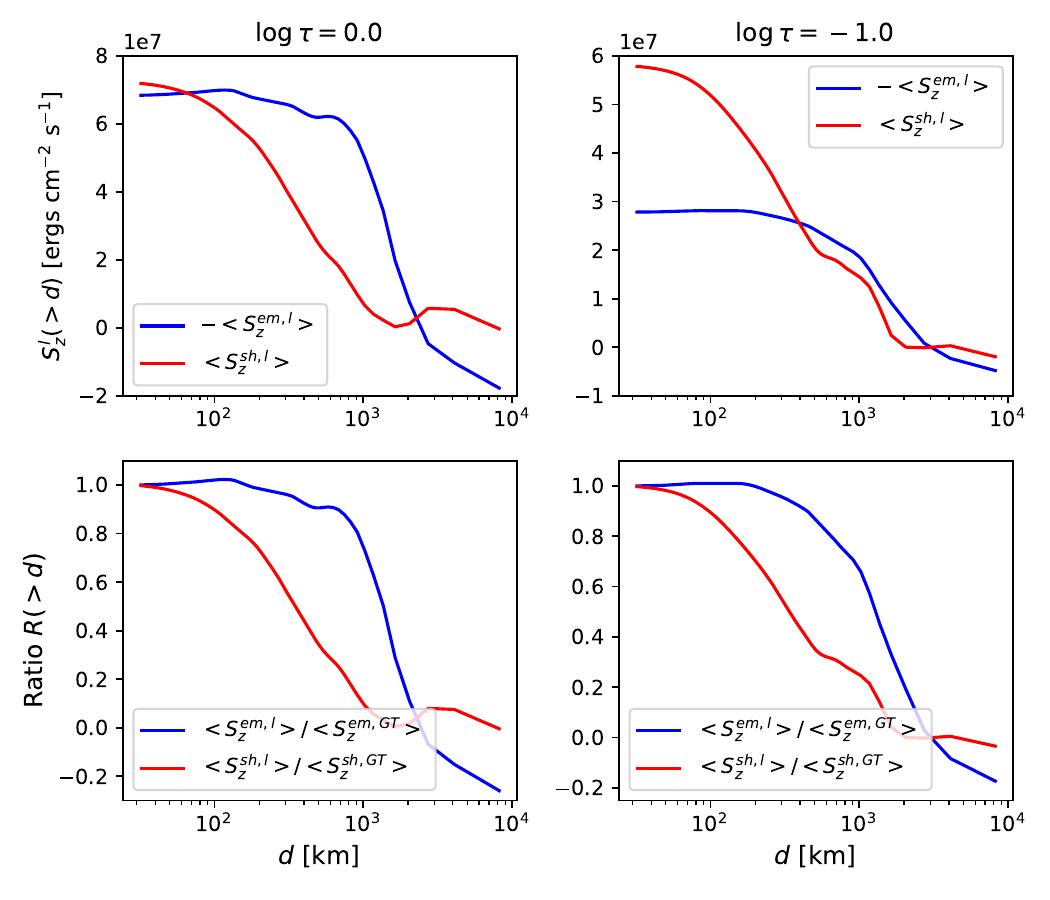}
  \caption{The response of emergence term $S_z^\text{em}$ and shearing term $S_z^\text{sh}$ of Poynting flux to the scale of velocity at $\log \tau = 0$ (left) and $\log \tau = -1$ (right). Top: the response of the absolute values to scale. Bottom: the response of the ratio of the value to ground truth. The red and blue lines represent the shearing term and emergence term, respectively.} 
  \label{fig:Poy_scale}
  \vspace{2mm}
\end{figure}

The relation between the two terms of $S_z^l$ is shown in Figure~\ref{fig:Poy_scale}. The upper panel shows the variation of the absolute values of $\left<S_z^\text{em}\right>$ and $\left<S_z^\text{sh}\right>$ with respect to scale of velocities $d$. The shearing term of Poynting flux has a steep slope: the small-scale velocity plays an important role. At $\log \tau = 0.0$, the two terms almost cancel out, and the net Poynting flux tends to be negative for large-scale velocity. The result at $\log \tau = -1.0$ differs from $\log \tau = 0.0$ mainly for small scales $d < 400$~km. The magnitude of the shearing term is about twice of the emergence term, which results in a clear net positive Poynting flux.

The bottom panel of Figure~\ref{fig:Poy_scale} shows the ratio of the $S_z^l$ to the ground truth $S_z$ with respect to the scale of velocity $d$. To recover $80\%$ of the shearing term of Poynting flux, the horizontal velocity $\bm{v}_h$ should have scales $d > 156$~km and $d > 145$ km at $\log \tau = 0$ and $-1.0$, while the vertical velocity $v_z$ with $d >  834$ km and $f > 625$ km can recover $80\%$ of the emergence term of Poynting flux at $\log \tau = 0$ and $-1.0$. At a scale of $d > 150$ km, the shear term is $80.8\%$ and $79.1\%$ of the ground truth value at $log \tau = 0.0$ and $-1.0$, respectively, and the emergence term is $100\%$ and $100\%$, respectively, which are close to the calculations of DAVE4VMwDV in Section \ref{sec:DAVE4VMwDV}.

We point out that the simulation output itself, which we use as the ground truth, may not strictly adhere to the ideal induction equation due to numerical diffusion. We compare $\partial B_z / \partial t$ and $\left(\bm{\nabla}\times(\bm{v}\times\bm{B})\right)_z$ on constant-$z$ slices. While the overall agreement is better than that for the for constant-$\tau$ slices show in the first row of Figure~\ref{fig:PSD}, the scatter becomes significantly greater below about $100$~km scale. This suggests that the numerical effect of the input likely also contributes to the discrepancy. For real-Sun observations, such non-ideal effect may come from small-scale, turbulent magnetic field and velocity, which manifests as a second order term in the induction equation that cannot be averaged to zero. It remains to be seen whether high-resolution observations from \textit{DKIST} can help alleviate the issue.

One interesting take-away from Figure~\ref{fig:Poy_scale} is that the induction equation (i.e., $\partial B_z/\partial t$ and $(\nabla \times \bm{E})_z$) is dominated by the spatial scale of a few hundred kilometers in this quiet-Sun simulation. As the geometric height of the input magnetograms (constant $\tau$) vary, one does not expect the induction equation to be well satisfied numerically beyond such size. Since DAVE4VMwDV solves for the induction equation ``locally'' by minimizing the residual within a small window, it might be more suited for quite-Sun calculations compared to other, more ``global'' algorithms.

\subsection{The Limitation in Stokes Inversion}

The second factor that can affect the estimate of Poynting flux is the magnetograms inverted with SIR. From Equation~(\ref{equ:Poynting}), the emergence term of the Poynting flux $\lvert S_z^\text{em} \rvert \propto B_h^2$ and the shearing term $\lvert S_z^\text{sh} \rvert \propto \lvert B_h B_z \rvert$. As demonstrated in Figure~\ref{fig:histogram}, the inaccuracies of the inferred magnetic field can affect the downstream velocity and the Poynting estimates.

One likely factor is the fact that we reduced the spatial resolution of the Stokes profiles before inversion, which is known to affect the retrieved magnetic field distribution \citep[e.g.,][]{milic2024}. The other factor is the inversion algorithm itself, for example, the node-based representation of the variables. As stated in Section~\ref{sec:Data}, to reduce the number of free parameters, SIR perturbed the initial atmosphere only in the location of nodes and approximates the remaining atmosphere with cubic spline interpolation. In our work, we apply five nodes in the atmosphere range in $\log \tau = [-4.2,1.0]$. The locations of nodes in the range $\log \tau = [-2.0,0]$(typical range of formation height for the \ion{Fe}{1} 630~nm lines) are $\log \tau = -0.3$ and $-1.6$. As the structure of the atmosphere is unknown \emph{a priori}, the ground-truth stratification may not be faithfully represented by a five-node function.

To investigate the effect of these factors, we conduct two more inversion experiments. For the first experiment, we keep the same configuration in Table \ref{tab:configuration} for SIR, but use synthetic Stokes without degradation. This experiment aims to assess the effects of PSF and rebinning. For the second experiment, we use the SIR configuration to those listed in Table \ref{tab:configuration_new} on the same input without degradation. We increase the number of nodes for magnetic field strength, inclination, and azimuth to $14$, $14$, and $5$. The locations of nodes for magnetic field strength and inclination in the range $\log \tau = [-2,0]$ are $[-0.2, -0.6, -1.0, -1.4, -1.8]$.

\begin{deluxetable}{ccccc}[t!]
  \tablecaption{Summary of SIR algorithm configuration for experiment two}\label{tab:configuration_new}
  \tablewidth{0pt}
  \tablehead{
  \colhead{}&
    \multicolumn{4}{c}{{Nodes}} \\
  \cline{2-5}
  \colhead{\textbf{Parameters}} & \colhead{\textbf{Cycle 1}} & \colhead{\textbf{Cycle 2}} & \colhead{\textbf{Cycle 3}} & \colhead{\textbf{Cycle 4}} 
  }
  \startdata
    Temperature     & 2 & 3 & 5 & 5 \\ 
    Microturbulence & 1 & 1 & 1 & 1 \\ 
    LOS velocity    & 1 & 2 & 3 & 5\\ 
    Magnetic field strength  & 1 & 2 & 3 & 14 \\ 
    Inclination     & 1 & 2 & 3 & 14 \\ 
    Azimuth         & 1 & 2 & 2 & 5 \\ 
  \enddata
  \end{deluxetable}
  \vspace{-5mm}

\begin{figure}[t!]
  \centering
  \includegraphics[width=0.45\textwidth]{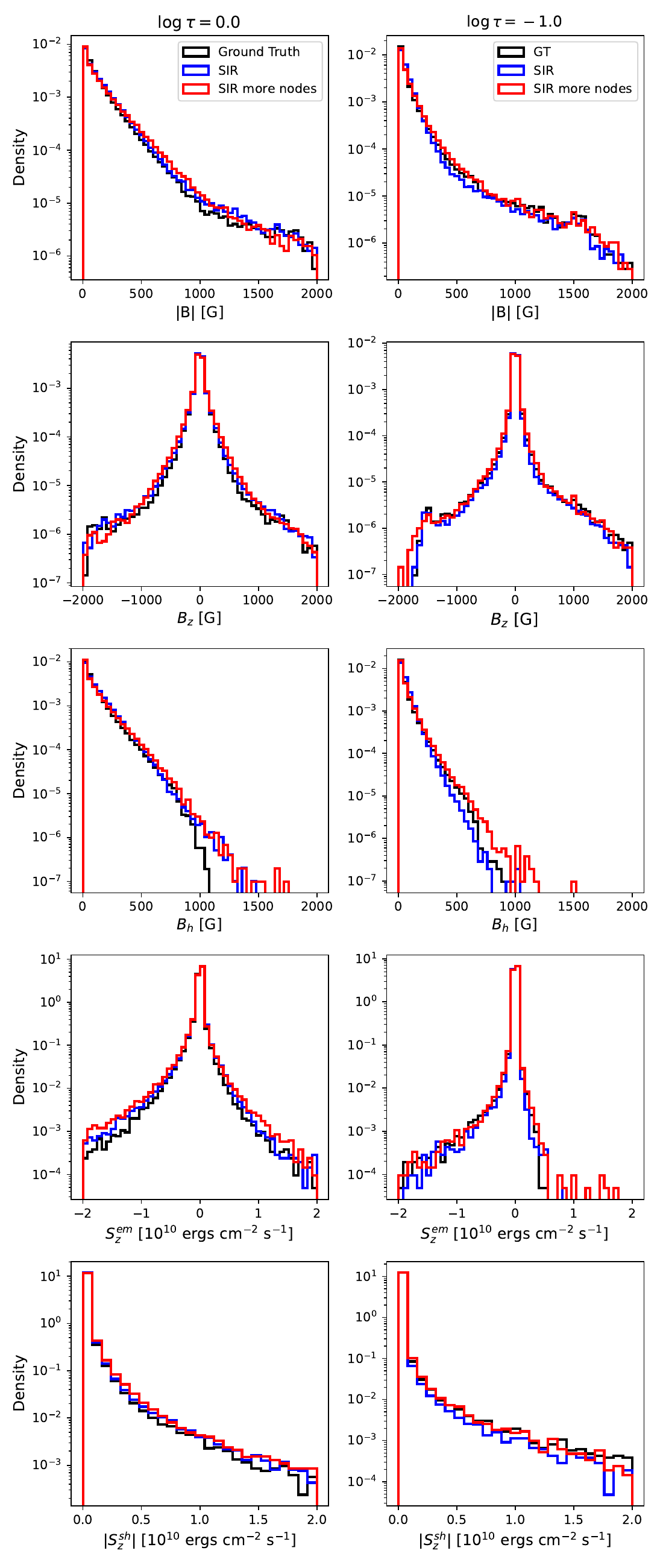}
  \caption{Comparison of the distributions of quantities from reference magnetic field (black), inverted magnetic field with configuration in Table \ref{tab:configuration} (red), and that with configuration in Table \ref{tab:configuration_new} (red) at $\log \tau = 0.0$ (left) and $\log \tau = -1.0$ (right). From top to bottom, we show the distribution of magnetic field strength $\lvert B \rvert$, vertical magnetic field $B_z$, horizontal magnetic field $B_h$, the emergence term of Poynting flux $S_z^\text{em}$, and unsigned shearing term of Poynting flux $S_z^\text{sh}$.} 
  \label{fig:histogram_all}
\end{figure}

The histograms of total, vertical, and horizontal magnetic fields from the two experiments, along with the reference values at optical depths $\log \tau = 0.0$ and $-1.0$, are shown in Figure \ref{fig:histogram_all}. For the first experiment (blue lines), the distributions of the inverted magnetic field still deviate from the reference value. But at $\log \tau =0$, the relative number of pixels with $\lvert B \rvert > 800$~{G} and $B_h > 750$~{G} is now greater than the reference, which is opposite to the result of inversion on degraded data. At $\log \tau = -1$, there are also deviation for $\lvert B \rvert$ in the range of $[250, 750]$~{G} and $B_h > 250$~{G}. The distribution of $B_z$ is smaller than the reference. The result of this experiment suggests that the degradation alone cannot fully account for the mismatch in the magnetic field distribution. The inversion process itself must also contribute.

For the experiment inversion with additional nodes, the distributions of inverted magnetic fields are greater than the reference at $\log \tau = 0.0$. At $\log \tau = -1.0$, the distributions of $B$ and $B_z$ are closer to the reference distributions. But the distribution of $B_h$ is greater than the reference at $B_h > 700$ G. This suggests that the fitting with more nodes may recover a more accurate distribution \citep{FIRTEZ-dz}. In practice, the optimal number of nodes cannot be determined a priori. Overfitting may be a concern.

\begin{figure}[t!]
  \centering
  \includegraphics[width=0.45\textwidth]{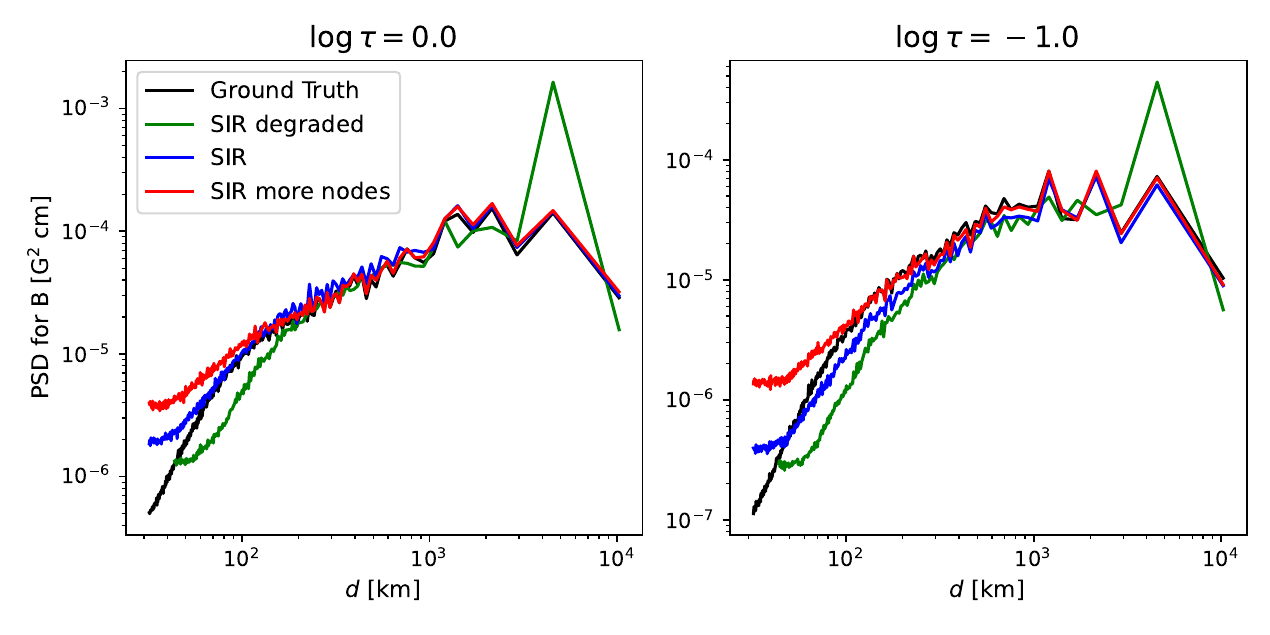}
  \caption{Comparison of the power spectrum density (PSD) of magnetic field strength from reference magnetic field (black), inversion from emulated observation (green), inverted magnetic field with configuration in Table \ref{tab:configuration} (red), and that with configuration in Table \ref{tab:configuration_new} (red) at $\log \tau = 0.0$ (left) and $\log \tau = -1.0$ (right).} 
  \label{fig:PSD_B}
\end{figure}

The PSDs of magnetic field strength for two experiments, inversion from emulated observation, and the reference are shown in Figure \ref{fig:PSD_B}. Compare to the reference, the overall quality of the inferred magnetic field strength is reasonable on larger spatial scales ($d\gtrsim200$~km), and the disagreement becomes obvious on smaller scales. Additionally, the result of the emulated observation displays an excess in PSD at the largest scale ($d\gtrsim4\times 10^3$~km), about half of the domain size. The reason for such an excess is unclear.

To investigate the effect of inversion on estimating Poynting flux, we calculate the Poynting flux with inferred magnetograms and ground truth velocities. Since there is still a problem in disambiguation, we only consider the emergence term ($S_z^{\text{em}}$) and the absolute value of shearing term ($\lvert S_z^{\text{sh}}\rvert$). The comparison between the two experiments and ground truth is shown in the last two rows of Figure \ref{fig:histogram_all}. At $\log \tau = 0.0$, as the result of greater relative numbers of large $B_z$ and $B_h$ in both experiments, the probability density distributions of $S_z^\text{em}$ and $\lvert S_z^\text{sh}\rvert$ of two experiments are greater than the reference, especially in the negative part of $S_z^\text{em}$. The results at $\log \tau = -1.0$ are different since the deviation in the distribution of magnetic field, $S_z^\text{em}$ and $\lvert S_z^\text{sh}\rvert$ both have smaller density than the reference for the first experiment. For the second experiment, the probability density distributions are close to reference for both $S_z^\text{em}$ and $\lvert S_z^\text{sh}\rvert$. The average emergence term of Poynting fluxes are $106.1\%$ and $147.0\%$ of the reference at $\log \tau = 0.0$, and $92.8\%$ and $99.3\%$ of the reference at $\log \tau = -1.0$ for two experiments, respectively. The average unsigned shearing term of Poynting fluxes are $116.9\%$ and $127.1\%$ of the reference at $\log \tau = 0.0$, and $78.3\%$ and $105.9\%$ of the reference at $\log \tau = -1.0$ for the two experiments, respectively. The results suggest that the missing magnetic flux and the overestimated magnetic flux from the inversion can result in missing Poynting flux and overestimating Poynting flux.

\subsection{Energy Transport in Quiet-Sun Photosphere}
\label{subsec:etqs}

\begin{figure}[t!]
  \centering
  \includegraphics[width=0.45\textwidth]{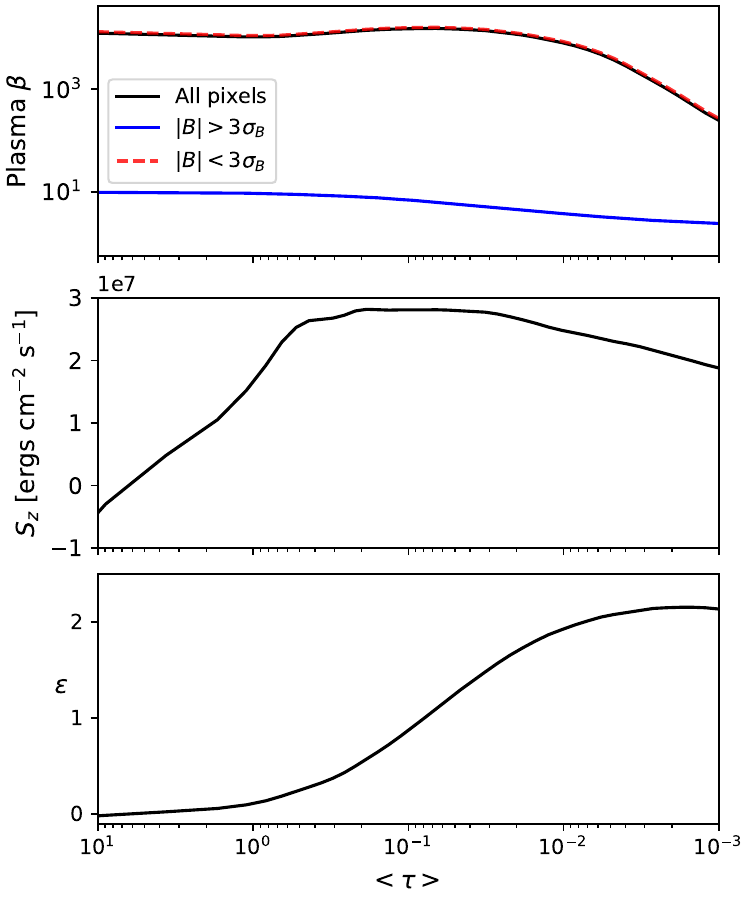}
  \caption{Top: Plasma $\beta$ in the region of interest with respect to height. The black, blue, and red lines represent the plasma $\beta$ averaged among all the pixels, pixels with $B < 3\sigma_{B}$ and pixels with $B > 3\sigma_{B}$ in the region of interest, respectively. Bottom: Fraction of energy transported by the Poynting flux relative to energy converted by the Lorentz force.} 
  \label{fig:beta}
  \vspace{2mm}
\end{figure}

Studying the energy transport in the quiet-Sun photosphere is difficult. Besides the difficulties in observation and inversion of Stokes profiles, another concern is that due to the weaker magnetic fields, the plasma $\beta=8 \pi p/B^2$ in the quiet-Sun photosphere varies more rapidly in space compared to active regions. The dynamic characteristics and the importance of the Poynting flux in energy transfer may change drastically with height. To demonstrate this, we plot the median plasma $\beta$ as a function of the logarithm median optical depth $\log \left<\tau\right>$ at each height in the top panel of Figure \ref{fig:beta}. The median plasma $\beta$ in the $\log \left<\tau\right>\in[0,-2]$ layer have a value of about $10^4$, suggesting that the plasma kinematics dominate over the magnetic fields. The behavior is quite different for regions with stronger (e.g., granular lanes) magnetic fields. Plasma $\beta$, where the magnetic field strength is greater than three times the standard deviation at each height ($B > 3\sigma_B$), has a value below $10$.

In the middle panel of Figure \ref{fig:beta}, we show the net Poynting flux density variation with height. The net Poynting flux increases with height in the range $\log \left<\tau\right> = [1, -0.5]$ and the value changes from $-4.0\times 10^6$~{ergs cm$^{-2}$ s$^{-1}$} to $2.6\times 10^7$~{ergs cm$^{-2}$ s$^{-1}$}. The net Poynting flux gradually decreases in the atmosphere higher than $\log \left<\tau\right> = -0.5$.
 
In the bottom panel of Figure \ref{fig:beta}, we calculate the contribution to energy change by Poynting flux. We calculate the ratio $\epsilon$ between the Poynting flux at a certain height $z$ and total magnetic energy change above height $z$, as a function of $\log \left<\tau\right>$ for each $z$:
\begin{equation}
  \epsilon = \frac{S_z(z)}{dE_m / dt},
\end{equation}
Here, the calculation is done on the entire box so the Poynting flux through the side boundary can be ignored. We also ignore the Poynting flux leaving the top boundary because its value is small \citep{Rempel2014}. At the height with $\log\left<\tau\right> = 0.0$, Poynting flux only contributes 11.2\% of the change of magnetic energy. This changes drastically at $\log \left<\tau\right> = -1.0$: Poynting flux contributes $87.3\%$ of the magnetic energy change, suggesting that it is the main source of magnetic energy change in the typically observed photospheric layer even under the non-force-free condition.

\subsection{Implications on Observations}

In a recent work, \cite{Tilipman} studied the Poynting flux in the quiet-Sun photosphere with real observation from the balloon-borne SUNRISE/IMaX data. They used two different codes based on deep learning and correlation tracking to infer the velocity and calculate the Poynting flux at $\log \tau \approx 0$. They found that there is not enough Poynting flux except for strong-field regions. Similar to our study on emulated observation, they also found that the shearing term of Poynting flux is underestimated. The emergence term dominates both total and unsigned Poynting flux, though the mean value was positive in their observation while it is negative in our study. The difference in sign may be attributed to the strong magnetic field in the upflow region of their ROI, while the emergence term of Poynting flux is concentrated on the intergranular lane. The fraction of emergence term on the unsigned Poynting flux is $99\%$ in their observation; the fraction is $92\%$ in our study. The small difference may be attributed to the use of different flow-tracking methods. The ability of difference flow tracking methods on estimating Poynting flux requires more investigation. 

The work of \cite{Tilipman} used a Milne-Eddington (ME) inversion code, which assumes constant magnetic field with height. Such a simplifying assumption is appropriate due to the limited spectral resolution of SUNRISE/IMaX. However, according to the MURaM simulation (Figure~\ref{fig:beta}), the Poynting flux in the quiet Sun has a large variation in height. A single Poynting flux estimate at $\log \tau \approx 0$ is not necessarily representative. 

\section{Conclusion}\label{sec:conclusion}

In this work, we assess the diagnostic capability of the Diffraction-Limited Near Infrared Spectropolarimeter (DL-NIRSP) instrument on \textit{DKIST} on the energy transport processes in the quiet-Sun photosphere with the realistic numerical MHD simulation code MURaM. We first validate the widely used and recently modified flow-tracking algorithm DAVE4VMwDV high-resolution data. We then synthesize and degrade high spatial and temporal resolution Stokes profiles of the \ion{Fe}{1} 630~nm lines to emulate the observation made by \textit{DKIST}/DL-NIRSP and infer the vector magnetic field maps. Finally, we examine the estimated Poynting flux from the emulated observation. 

Our main findings are as follows:
\begin{enumerate}
  \item DAVE4VMwDV works reasonably well on inferring the photospheric velocity on high-resolution data, especially the large-scale ($d > 150$ km) velocity. At smaller scales, the corrugated $\tau$ surface results in the induction equation being poorly satisfied via numerical differentiation; the accuracy of the velocity inference thus suffers. For the emulated observation, the ill-satisfied induction equation does not impact the large-scale velocity much.
  \item The temporal resolution of input magnetograms can affect the performance of DAVE4VMwDV. To get a reasonable velocity estimate, we suggest the input magnetograms and Dopplergrams have $\Delta x / \Delta t > \left<v\right>$. 
  \item The Poynting flux calculated with inferred velocity similarly matches the ground truth better at larger spatial scales. The averaged value is underestimated in both simulation and emulated observation. On the emulated observation, the estimated unsigned Poynting flux is about $72.5\%$ and $61.3\%$ of the reference ground truth at $\log \tau = 0.0$ and $\log \tau = -1.0$. The net Poynting flux is $-1.3\times 10^7$ ergs cm$^{-2}$ s$^{-1}$ and $-8.5\times 10^6$ ergs cm$^{-2}$ s$^{-1}$, compared to ground truth $3.7\times 10^6$ ergs cm$^{-2}$ s$^{-1}$ and $3.0\times 10^7$ ergs cm$^{-2}$ s$^{-1}$. 
  
  \item The main difference comes from the underestimated shearing term. The estimated net emerging Poynting flux can reproduce $99.6\%$ and $88.8\%$ of the ground truth net emerging Poynting flux, while the net shearing Poynting flux is only $72.0\%$ and $55.5\%$ of the ground truth. 

  \item The net shearing term of the Poynting flux is more difficult to estimate than the net emergence term of the Poynting flux. To recover $80\%$ of the shearing term of the Poynting flux, the inferred horizontal velocity field should have a spatial scale $d > 156$ km and $d > 145$ km at $\log \tau = 0$ and $-1$. To recover $80\%$ of the emergence term of the Poynting flux, the inferred vertical velocity field should have a spatial scale $d > 834$ km and $d > 625$ km.
\end{enumerate}

Estimating energy transport on the photosphere of the quiet Sun is difficult. \chg{The inference of the magnetic field and the velocity field must both be improved in order for a better estimate of the Poynting flux. For magnetic field inversion, new advances have been made by including the more realistic \textit{magnetohydrostatic} equilibrium assumption \citep[compared to the commonly used hydrostatic equilibrium;][]{FIRTEZ-dz}, or using deep-learning algorithms \citep{ML_2019,SPIN4D}. These methods can resolve azimuthal ambiguity, and provide an absolute spatial scale along the LOS, which produces magnetic field maps on a constant geometric height. The latter makes the application of velocity tracking an self-consistent approach, and can also reduce the effect of the $p$-mode that would imprint on the $\tau$ surface (M. Rempel, private communication). To demonstrate the possible improvement, we applied DAVE4VMwDV on the MURaM simulation at a two constant geometric heights with $\left\langle\log\tau\right\rangle = 0.02$ and $\left\langle\log \tau \right\rangle  = -1.02$. At these two heights, 82.9\% and 74.5\% of the shearing term can be recovered, which shows about 10\% and 20\% increase compared to the results for the constant-$\tau$ surfaces.}

\chg{For velocity inference, new algorithms based on deep learning, such as DeepVel \citep{Deepvel}, have shown much potential. High spectral and spatial resolution observations from \textit{DKIST} and multi-spectral-line Stokes inversions will also allow for the estimates of the velocity field or electric field at multiple heights simultaneously. This is expected to improve the study on the energy transport in the solar atmosphere above the quiet Sun.}

\chg{In addition, the signal-to-noise level is another important factor in estimating the Poynting flux. \cite{quiteronoda2023} studied the effect of noise on Stokes inversion. Their study suggested that the inversion of data with noise level on the order of or lower than $5\times 10^{-4}$ of continuum intensity can provide more reliable information for the vector magnetic fields. The detail of how noise affects the Poynting flux estimate will be investigated in the future.}



\begin*{Acknowledgments}
We thank Carlos Quintero Noda and Matthias Rempel for their valueable input. Support for this work is provided by the National Science Foundation through the \textit{DKIST} Ambassadors program, administered by the National Solar Observatory and the Association of Universities for Research in Astronomy, Inc. X. S. is additionally supported by NSF Awards \#2008344 and \#1848250.
\end*{Acknowledgments}


\begin{figure}[th!]
  \centering
  \includegraphics[width=0.45\textwidth]{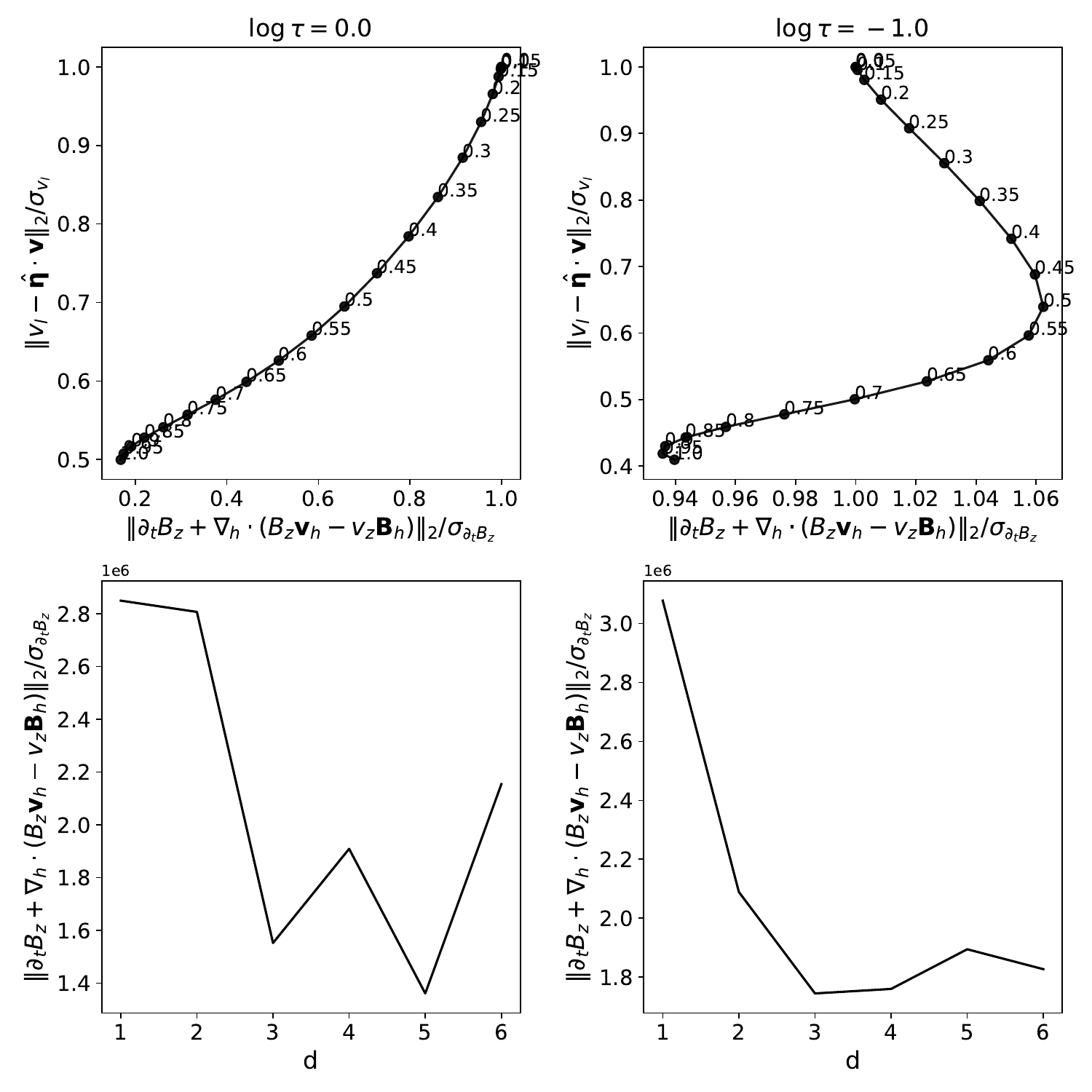}
  \caption{Parameter selection for DAVE4VMwDV for MHD data at $\log \tau = 0$ (left) and $\log \tau = -1$ (right). Top: L-curve displaying the tradeoff between the two loss function terms $L_1$ and $L_2$ in DAVE4VMwDV. The values next to the black dots represent the weighting $\lambda$ of each test. Bottom: The variation of loss function terms $L_1$ with respect to degree of Legendre expansion in horizontal direction $d$.} 
  \label{fig:L_curve_MHD}
  \vspace{2mm}
\end{figure}

\appendix 

\section{Optimization of DAVE4VMwDV}\label{app:optimization}

\chg{Besides the magnetogram and Dopplergram input, DAVE4VMwDV has three free parameters: the window size $w$ for optimization, the maximum degree of Legendre polynomials $d$ for velocity expansion inside the window, and the relative weighting $\lambda$ for term $L_2$ defined in Equation set \ref{l2}. In this work, we optimize these parameters using the methods described in \cite{Liu_optimize}. }The initial parameters are $w = 15$, $d = 3$, and $d_r = 5$.

\subsection{DAVE4VMwDV on MHD simulation}\label{app:opt_MHD}

\chg{We first use ``L-curve'' that displays the trade-off between the two terms as $\lambda$ varies \citep[e.g.,][]{Lcurve} to determine the $\lambda$. Figure~\ref{fig:L_curve_MHD} shows curves of normalized $(L_1,L_2)$ with varied $\lambda$ at $\log \tau = 0$ and $\log \tau = -1$ for the application to the MHD data. The point $\lambda = 0$ represents the case of DAVE4VM, where the Doppler constraint is not considered.} At $\log \tau = 0.0$, both $L_1$ and $L_2$ decrease as $\lambda$ increases, while at $\log \tau = -1.0$, the curve likes an inverse ``L'' and both $L_1$ and $L_2$ are small at large $\lambda$. Therefore we choose $\lambda = 1.0$.


We then choose the degree of Legendre expansion in the horizontal direction $d$ by comparing the value of $L_1$ from different $d$. The variation of $L_1$ with respect to the degree of Legendre expansion in the horizontal direction $d$ at two layers is shown in the bottom panel of Figure~\ref{fig:L_curve_MHD}. The degree of Legendre expansion in the vertical direction $d_r$ is set as $d_r = d + 2$. At these two layers, we choose $d = 5$ and $d = 3$ because they have the smallest $L_1$.

\subsection{DAVE4VMwDV on emulated observation}\label{app:opt_obs}
\begin{figure}[t!]
  \centering
  \includegraphics[width=0.48\textwidth]{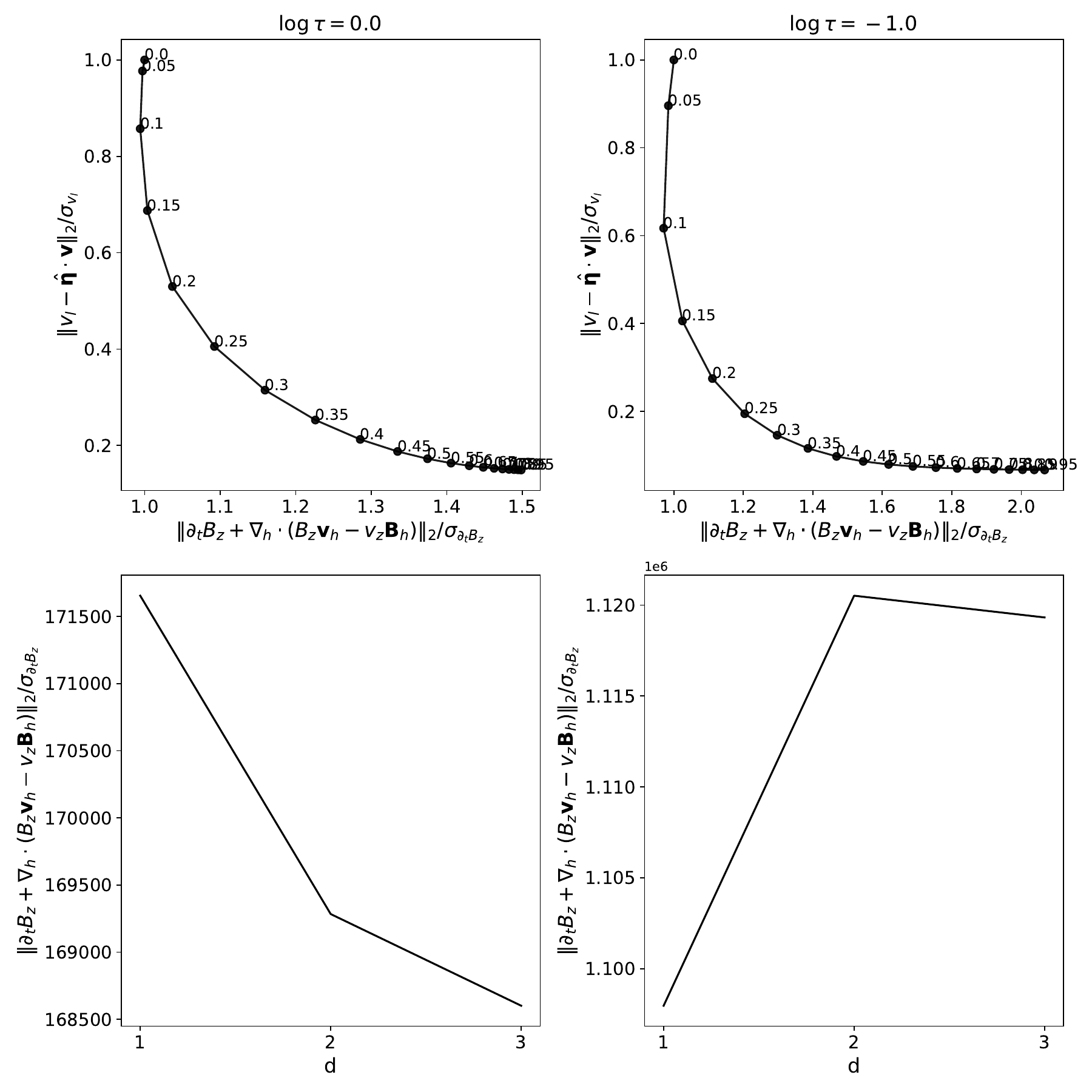}
  \caption{Similar to Figure \ref{fig:L_curve_MHD}, but for emulated observation} 
  \label{fig:L_curve}
  \vspace{2mm}
\end{figure}

Figure~\ref{fig:L_curve} shows the L-curves with $\lambda$ varies from $0$ to $1$ for the application on emulated observation at $\log \tau = 0$ and $\log \tau = -1$.

The L-curves for $\log \tau = 0$ and $\log \tau = -1$ have similar shape: \chg{as the weighting $\lambda$ increases, the Doppler residual $L_2$ decreases drastically first and then decreases at a slower pace, while the induction term $L_1$ has the inverse trend.} The $\lambda=0.35$ is finally selected for both heights as it provides a reasonable compromise: $L_2$ drops by factors of 3 and 5 compared to DAVE4VM, while the increases in $L_1$ are only 25\% and 35\%, respectively.

The variation of $L_1$ with respect to degree of Legendre expansion in the horizontal direction $d$ at two layers are shown in the bottom panel of Figure~\ref{fig:L_curve}. The degree of Legendre expansion in horizontal direction $d_r$ is set to $5$. The variation of $L_1$ for $d = [1, 2, 3]$ at $\log\tau =0$ and $\log\tau =-1$ are plotted in Figure \ref{fig:L_curve}.

As shown in the left panel, at $\log\tau = 0.0$, $L_1$ decreases with respect to $d$ and $d = 3$ gives a smallest value for $L_1$, suggesting the best fit for induction equation. However, in the case of $\log\tau = -1.0$ (right panel), $L_1$ shows a different variation to the case of $\log\tau = 0.0$. In this case, $d = 1$  gives the best fit to induction equation. 

\section{DAVE4VMwDV on Successive Frames}\label{app:successive}

In the literature, the velocity is often estimated from two successive frames of observation to maximize the output cadence. For two frames at $t_0$ and $t_1 = t_0+\Delta t$, where $\Delta t$ is the cadence of the observation, the output is designated to the averaged time stamp, $t=t_0+\Delta t/2$. 
The time derivative of a function $f$ is calculated as 
\begin{equation}
  \begin{split}
    \left.\frac{\partial f}{\partial t}\right|_{t = t_0+\frac{\Delta t}{2}} &= \frac{f(t_0+\Delta t) - f(t_0)}{\Delta t}. \\  
  \label{forward}
  \end{split}
\end{equation}
The input vector magnetograms and Doppler velocity are the average of two frames 
\begin{equation}
  f(t_0+\frac{\Delta t}{2}) = \frac{f(t_0+\Delta t) + f(t_0)}{2},
\end{equation}
where $f$ can represent $B_x$, $B_y$, $B_z$, or $v_l$.

We posited earlier that the poor performance of DAVE4VMwDV might be due to the violation of the CFL condition. Our choice of temporal difference limit the time difference of the input at $4$~s, twice of that of the MURaM simulation snapshots. In this section, we apply DAVE4VMwDV on MURaM data at $t_0 = 32$~s and $t_1 = 34$~s, i.e., the highest possible cadence, to estimate the velocity at $t = 33$~s. For Poynting flux calculation, the other inputs are the averaged magnetogram and Dopplergram of these two time steps. We quantify the overall performance by evaluating the coefficients listed in Section \ref{subsec:Performance}. The ground truth velocity is the averaged velocity from the simulation at $t_0$ and $t_1$.

\begin{figure*}[t!]
  \centering
  \includegraphics[width=0.95\textwidth]{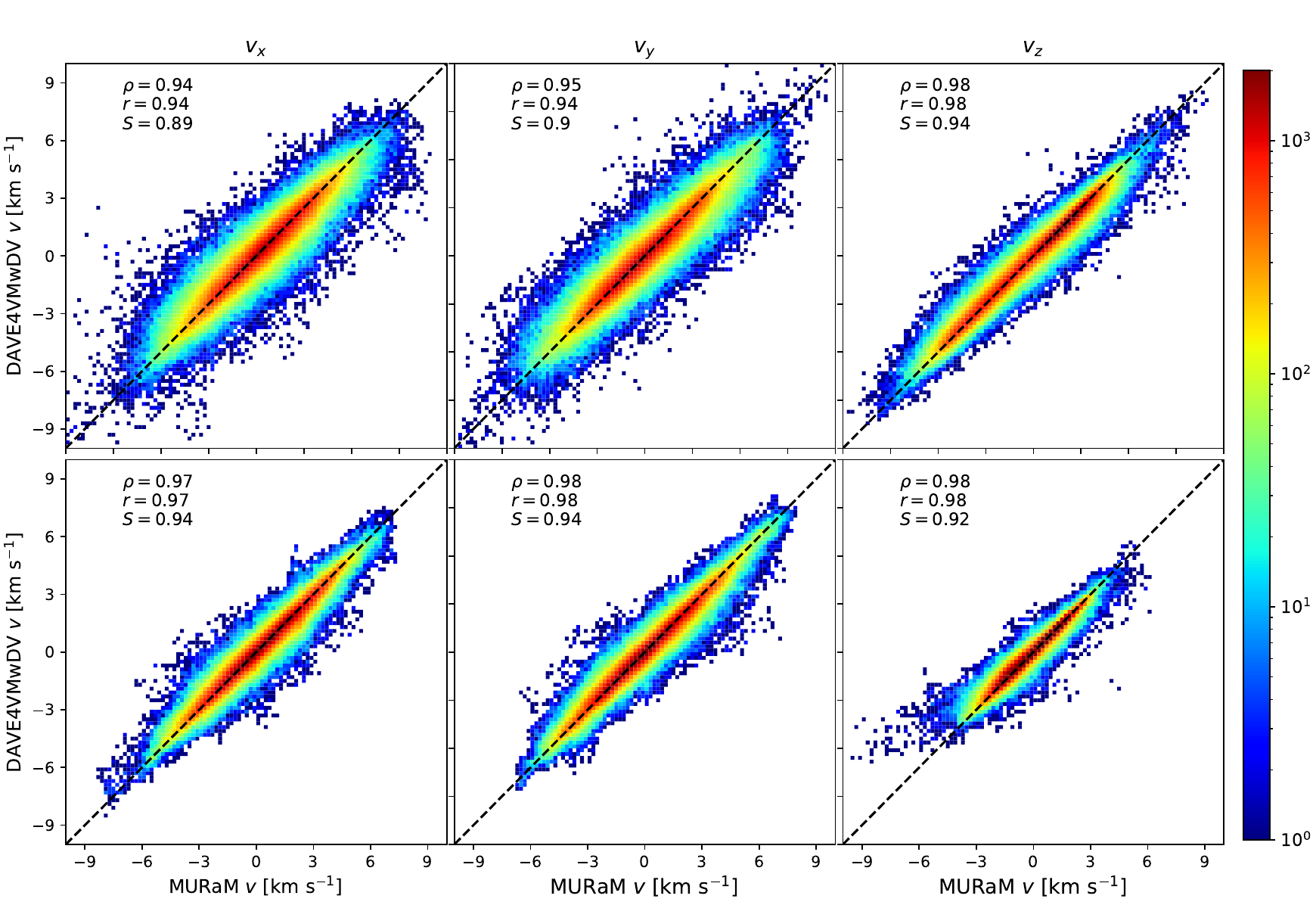}
  \caption{2D histograms of the inferred velocity field and the reference velocity field at $\log \tau = 0$ (top) and $\log \tau = -1$ (bottom). From left to right, we show the histograms for $v_x$, $v_y$, and $v_z$, respectively. The Spearman coefficient ($\rho$), Pearson coefficient ($r$), and slope ($S$) are also shown on the plots.}
  \label{fig:vel_hist_mid_point}
  \vspace{2mm}
\end{figure*}

Figure~\ref{fig:vel_hist_mid_point} shows the scatter plots between inferred and the ground-truth velocities at two optical depths. The inferred velocity at $\log \tau = 0$ has $E_{\text{rel}} = 0.28$, $C = 0.96$, and $A = 0.95$, while the inferred velocity at $\log \tau = -1$ has $E_{\text{rel}} = 0.23$, $C = 0.97$, and $A = 0.97$. Compared to the case with $\Delta t = 2$~s in Section \ref{subsec:Performance}, the performance is slightly increased. The slopes are closer to $1$ for $v_x$ and $v_y$. Similar to the findings in Section \ref{subsec:Poynting}, the net shearing term of the Poynting flux is underestimated for both layers. The estimated net shearing Poynting flux can reproduce $70.0\%$ and $57.8\%$ of the ground truth at $\log \tau = 0$ and $\log \tau = -1$, respectively. The estimated net emergence Poynting flux can reproduce $98.5\%$ and $90.4\%$ of the ground truth at $\log \tau = 0$ and $\log \tau = -1$, respectively. 


\bibliography{reference}{}
\bibliographystyle{aasjournal}


\end{CJK*}
\end{document}